\documentclass[12pt]{article}
\usepackage{geometry}
\usepackage{amsmath}
\usepackage{amssymb,epsfig,subfigure}
\usepackage{graphicx}
\numberwithin{equation}{section}

\newcommand{\be}{\begin{equation}}
\newcommand{\ee}{\end{equation}}
\newcommand{\ba}{\begin{aligned}}
\newcommand{\ea}{\end{aligned}}
\newcommand{\half}{{1\over 2}}

\newcommand{\E}{\mathcal{E}}
\newcommand{\talpha}{{\tilde \alpha}}
\newcommand{\tbeta}{{\tilde \beta}}
\newcommand{\tgamma}{{\tilde \gamma}}
\newcommand{\tdelta}{{\tilde \delta}}

\def\m1{\left(-1\right)^{F_i}}
\newcommand{\dg}{\dagger}
\makeatletter
\def\sla@#1#2#3#4#5{{%
  \setbox\z@\hbox{$\m@th#4#5$}%
  \setbox\tw@\hbox{$\m@th#4#1$}%
  \dimen4\wd\ifdim\wd\z@<\wd\tw@\tw@\else\z@\fi
  \dimen@\ht\tw@
  \advance\dimen@-\dp\tw@
  \advance\dimen@-\ht\z@
  \advance\dimen@\dp\z@
  \divide\dimen@\tw@
  \advance\dimen@-#3\ht\tw@
  \advance\dimen@-#3\dp\tw@
  \dimen@ii#2\wd\z@  \raise-\dimen@\hbox to\dimen4{%
    \hss\kern\dimen@ii\box\tw@\kern-\dimen@ii\hss}%
  \llap{\hbox to\dimen4{\hss\box\z@\hss}}}}
\def\slashed#1{%
  \expandafter\ifx\csname sla@\string#1\endcsname\relax
    {\mathpalette{\sla@/00}{#1}}%
  \else
    \csname sla@\string#1\endcsname
  \fi}
\makeatother

\begin{document}


\thispagestyle{empty}
\begin{flushright}\footnotesize
\texttt{arXiv:0808.1286}\\
\texttt{CALT-68-2697}\\
\vspace{1.8cm}
\end{flushright}

\renewcommand{\thefootnote}{\fnsymbol{footnote}}
\setcounter{footnote}{0}

\begin{center}
{\Large\textbf{\mathversion{bold} Instantons and SUSY breaking
 in F-theory}\par}

\vspace{2.1cm}

\textrm{Jonathan J. Heckman$\,^{\sharp}$, Joseph Marsano$\,^{\natural}$, Natalia Saulina$\,^{\natural}$,}
\vspace{0.3cm}

\textrm{Sakura Sch\"afer-Nameki$\,^{\natural}$ and Cumrun Vafa$\,^{\sharp}$}

\vspace{1cm}

\textit{$\,^{\sharp}$ Jefferson Physical Laboratory, Harvard University,
Cambridge, MA 02138, USA}\\
\texttt{jheckman@fas.harvard.edu, vafa@physics.harvard.edu}

\vspace{.6cm}

\textit{$\,^{\natural}$California Institute of Technology\\
1200 E California Blvd., Pasadena, CA 91125, USA } \\
\texttt{marsano, saulina, ss299 @theory.caltech.edu}

\bigskip


\par\vspace{1cm}

\textbf{Abstract}\vspace{5mm}
\end{center}

\noindent
We study instanton contributions to the superpotential of local F-theory compactifications
which could potentially be used to engineer models of dynamical supersymmetry breaking.
These instantons correspond to Euclidean 3-branes which form a threshold bound
state with spacetime filling 7-branes.  In certain cases, their contributions to the effective 4d superpotential
can be determined in both perturbative string theory
as well as directly via the topologically twisted theory on the 3-brane worldvolume,
and in all cases we observe an exact match between these results.
We further present an instanton generated Polonyi-like model, and characterize
subleading corrections to the superpotential which arise from multi-instantons.
We also study instanton contributions to 4d pure $\mathcal{N}=1$
$SU(N)$ gauge theory realized by a stack of 7-branes wrapping a rigid 4-cycle and find that there is a non-trivial contribution
to the glueball superpotential from the single instanton sector.  This correction is absent in the purely 4d theory
and could conceivably be used either to stabilize moduli or to break supersymmetry.

\vspace*{\fill}

\setcounter{page}{1}
\renewcommand{\thefootnote}{\arabic{footnote}}
\setcounter{footnote}{0}

 \newpage


\tableofcontents


\section{Introduction}

Any semi-realistic string construction of the minimal supersymmetric standard model (MSSM) must eventually
address the origin of supersymmetry (SUSY) breaking.
Determining the dynamics of the hidden sector and how it
couples to the visible sector has the potential to
shed important light on both the low energy dynamics of the
MSSM as well as the high energy behavior of any candidate string compactification.

A priori, the hidden sector could potentially be highly stringy in origin or
could reduce to conventional field theory dynamics.  Metastable brane/anti-brane
configurations provide a tractable example of the former possibility
\cite{Aganagic:2006ex, Heckman:2007wk, Marsano:2007fe, Heckman:2007ub}. D-brane probes of
local geometric singularities reduce at low energies to quiver gauge theories, and in
appropriate circumstances can break supersymmetry due to strongly coupled field theory dynamics
\cite{Berenstein:2005xa, Franco:2005zu, Bertolini:2005di, Diaconescu:2005pc, Florea:2006si}.
It has so far proved elusive, however, to construct string vacua which combine such candidate
hidden sectors with a visible sector containing the MSSM.

Given the fact that the visible sector of any string construction is likely
to be quite complicated, at a pragmatic level it is natural to
simplify as much as possible the content of the hidden sector.  Retro-fitted
examples of the classic Polonyi, Fayet and O'Raifeartaigh models of supersymmetry
breaking were studied in \cite{Dine:2006gm} and have subsequently been constructed in
local IIB setups \cite{Aharony:2007db, Aganagic:2007py}. In these constructions,
D5-branes wrap various $\mathbb{P}^1$'s of a local Calabi-Yau 3-fold and supersymmetry is
broken by D1-instantons wrapping these same $\mathbb{P}^1$'s.  Of particular importance
is the fact that the scale of supersymmetry breaking is exponentially suppressed by
the D1-instanton action. Similar Polonyi models have also been realized in compact
IIB orientifold constructions \cite{Buican:2008qe}, as well as in compactifications
of Type I string theory \cite{Cvetic:2007qj,Cvetic:2008mh}.

In the context of F-theory compactified on a local Calabi-Yau 4-fold, the search for a viable
visible sector can also be significantly constrained by assuming that a supersymmetric
grand unified theory (GUT) exists at high energies, and that this GUT admits a decoupling limit
\cite{Beasley:2008dc, Beasley:2008kw}.  See also \cite{Donagi:2008ca, Hayashi:2008ba, Aparicio:2008wh} for related investigations and \cite{Buchbinder:2008at} for a study of ISS SUSY-breaking in this context.  In these local models, the gauge degrees of freedom of the GUT descend from the
worldvolume of a spacetime filling 7-brane wrapping a compact K\"ahler surface, and the chiral matter of the MSSM localizes
on Riemann surfaces where the GUT model 7-brane intersects additional 7-branes.  As shown in \cite{Beasley:2008kw},
simply achieving the correct chiral matter content of the MSSM through an appropriate choice of background fluxes automatically
addresses several puzzles of traditional 4d GUT models.  Although quite encouraging, these results
were limited to supersymmetric models.  To make further contact with observation, it is necessary to couple this class
of models to a hidden sector which breaks supersymmetry.

The goal of this paper is to take a first step in this direction by studying how stringy instantons,
which can be viewed as instantons in a higher dimensional gauge theory,
can generate non-trivial hierarchically suppressed contributions to the 4d superpotential in F-theory of the sort that can be used to dynamically break supersymmetry at a naturally small scale.  We will further present explicit examples of such superpotential contributions.  This includes a Polonyi-like superpotential, and a nontrivial superpotential correction to pure ${\cal{N}}=1$ $SU(N)$ gauge theory engineered from 7-branes wrapping a rigid 4-cycle, which has potential applications to moduli stabilization and possibly supersymmetry breaking as well.  Ultimately, these ideas can be used to implement explicit models of SUSY breaking in F-theory GUT models, and two distinct applications along these lines will appear in \cite{HV} and \cite{MSSN}.


A study of possible non-perturbative contributions to the effective superpotential in F-theory
was initiated in \cite{Witten:1996bn} and has been further analyzed, for
example, in \cite{Ganor:1996pe}. The case of instanton corrections in the presence of background
fluxes is considerably more subtle, and only partial results are available
\cite{Robbins:2004hx, Saulina:2005ve, Kallosh:2005gs, Lust:2005cu, Tsimpis:2007sx}.  Indeed, when D3-instantons
wrap a surface of general type, there will typically be additional zero modes which must be lifted via fluxes
in order to obtain a non-trivial contribution to the superpotential.  To bypass this issue, we shall therefore
restrict attention to D3-instantons wrapping del Pezzo surfaces, where such subtleties are absent.

We find that in appropriate circumstances D3-instantons will contribute
to the superpotential when they wrap the same 4-cycle wrapped by a 7-brane.  Contrary to what seems to be standard lore, this includes some situations
where the D3-instanton worldvolume theory admits a non-trivial background
flux different from that already present on the 7-brane. As we explain in Section \ref{sec:generalities}, at a heuristic level this
is consistent with the arithmetic genus condition on possible M5-instantons in the dual M-theory description \cite{Witten:1996bn}.

The explicit example of a Polonyi-like model that we find originates from a configuration of
two spacetime filling 7-branes which wrap distinct del Pezzo surfaces and intersect along a matter curve.
In our construction, each 7-brane has a $U(1)$ gauge symmetry. For an appropriate choice of internal
fluxes on these 7-branes, the low energy theory will contain a single 4d chiral superfield, $X$.
Summing over all relevant $U(1)$ fluxes in the internal worldvolume theory of a single
D3-instanton which wraps one of these del Pezzo surfaces, we show that the 1-instanton sector
generates a linear superpotential term $W_{\text{inst}}=F_XX$ and no higher powers of the form
$X^m$ for $m>1$ so that, to leading order, this theory behaves as a Polonyi model.

Beyond the 1-instanton sector, the leading order behavior of the superpotential
receives additional, exponentially suppressed corrections. The multi-instanton sector
of the theory corresponds to configurations where multiple D3-instantons wrap surfaces in the geometry.
We study the worldvolume theory of a stack of multiple D3-instantons wrapping one of the del Pezzo
surfaces of the model and formulate conditions for the internal fluxes in this configuration to
generate higher order contributions to the superpotential of the form $X^m$ with $m\ge 1$.
We present some examples which show that such contributions are indeed present, though they are highly suppressed near the origin $X=0$.
Once a particular K\"ahler potential has been specified, this type of superpotential can be used to
engineer an explicit model of supersymmetry breaking.  Although a full study of the K\"ahler potential is
beyond the scope of this paper, we note that in similar purely 4d Polonyi models with an anomalous $U(1)$ gauge symmetry,
integrating out the associated massive gauge boson generates a correction to the K\"ahler potential for the $X$ field which
has the correct sign to generate a supersymmetry breaking minimum \cite{ArkaniHamed:1998nu}.

We also find that D3-instantons produce important corrections to the low energy dynamics of pure $\mathcal{N} = 1$ $SU(N)$ gauge theory
engineered from a stack of 7-branes wrapping a del Pezzo surface in F-theory.  In particular, we determine a 1-instanton correction
to the Veneziano-Yankielowicz \cite{Veneziano:1982ah} glueball superpotential which is absent in the purely 4d theory.
Even in a local model, the volume of the del Pezzo surface is a dynamical modulus, and it is therefore
appropriate to study the effective potential defined by the glueball field and volume
modulus.  This is to be contrasted with the case of pure $\mathcal{N} = 1$ $SU(N)$ gauge theory engineered from D5-branes
wrapping a rigid $\mathbb{P}^{1}$ in a local Calabi-Yau 3-fold as in \cite{Vafa:2000wi, Cachazo:2001jy}, where the corresponding
volume modulus is non-dynamical. Along these lines, we briefly comment on potential applications to moduli stabilization as
well as supersymmetry breaking.

The organization of this paper is as follows.  In Section \ref{sec:generalities} we describe in general terms our methodology for
computing the zero mode content of D3-instantons in F-theory.  Additional details on this procedure can be found in Appendix B.
In Section \ref{sec:polonyi} we present our Polonyi-like model, and in Section \ref{sec:puresun} we study the one instanton sector of pure 4d $\mathcal{N} = 1$
$SU(N)$ gauge theory engineered by a 7-brane wrapping a rigid 4-cycle.

\vspace{.5cm}

\section{D3-Instantons -- Generalities}
\label{sec:generalities}

Before proceeding to any specific model, in this Section we show from complementary
viewpoints that in compactifications of F-theory on a Calabi-Yau 4-fold, D3-instantons
can contribute to the superpotential of the low energy theory defined by one or
more 7-branes wrapping a rigid K\"ahler surface $S$ in the 3-fold base $B$ of the $CY_{4}$.
The long wavelength limit of the theory defined by a 7-brane wrapping $S$ is specified by a unique
topological twist \cite{Beasley:2008dc} which we briefly review in Appendix B.  In particular, because the
Euclidean 3-brane wraps the same K\"ahler surface, there is a unique twist for this worldvolume theory as well,
and it is given by dimensional reduction of the 7-brane result.  The collective coordinates of the Euclidean 3-brane configuration
can then in principle be determined using the same procedure for zero mode counting developed for the 7-brane theory.

In certain cases, Euclidean 3-branes wrapping the same surface $S$ admit
an interpretation in terms of ordinary gauge theory instantons. This is the
case, for example, when the net flux through both the 7-brane and 3-brane is trivial.  The net instanton contribution
from such 3-branes is given by summing over all consistent internal flux sectors{\footnote{A given flux is ``consistent'' if it does not lift the collective coordinates of the D3-brane.  As we shall see later, this corresponds to fluxes that are ``trivializable'' in the sense that they are dual inside $S$ to 2-cycles that are trivial in $B$.}} of the 3-brane worldvolume theory.
An important consequence of this fact is that such instantons can potentially generate superpotential terms in 4d $U(1)$ gauge theories.
While this is a perhaps more exotic possibility from the purely 4d viewpoint, from the perspective of the 7-brane theory, such
contributions will typically be present.  Indeed, an internal flux through the
7-brane will induce lower brane charges determined by the appropriate Chern character of the background gauge bundle.
This intuition serves as a useful check on the explicit calculations we shall perform.

Now, although F-theory may be viewed as a non-perturbative extension of type IIB string theory,
when the bound state of 7-branes wrapping a surface $S$ are all of the same $(p,q)$ type, there exists an
element of the duality group $SL(2,\mathbb{Z})$ which transforms this into a configuration of D7-branes.  In particular,
because D3-branes remain invariant under the entire symmetry group, we conclude that in many cases it is appropriate
to analyze the resulting configuration using perturbative string theory.  The zero mode content is then given by appropriate
$3$-$3$, $3$-$7$ and $7$-$3$ strings.  When 7'-branes of the same $(p,q)$ type intersect the 7-brane wrapping $S$, $3$-$7'$ and $7'$-$3$
strings can also contribute perturbatively.

The existence of a perturbative limit also provides a useful check on the results of the topological
field theory analysis, and in all cases where a direct comparison is available, we obtain an exact
match with the perturbative result.
The only significant difference between the two approaches, in fact,
is that in the former case, the fields are twisted by the canonical bundle of the
surface, $K_{S}$, whereas in the perturbative case, the fields are twisted by the normal bundle
$\mathcal{N}_{S/B}$ for $S$ in a Calabi-Yau 3-fold.  Although in F-theory the 3-fold base may
not satisfy the Calabi-Yau condition, this distinction is largely unimportant because there
is a unique twist available which is compatible with $\mathcal{N} = 1$ supersymmetry in four dimensions.

In fact, for the most part, the moduli space of the 7-brane
behaves as if the 7-brane were embedded in a Calabi-Yau threefold.  Indeed, the total space for the $(2,0)$ form
moduli space is $\mathcal{O}_{S}(K_{S}) \rightarrow S$, which is a non-compact Calabi-Yau 3-fold. As emphasized
in \cite{Donagi:2008ca}, when a heterotic dual is available, this Calabi-Yau 3-fold corresponds to a non-compact
limit of the corresponding heterotic compactification. In all of this paper, the singularity type of the elliptic
fibration is simple enough that such subtleties can for the most part be ignored.
In order to more readily check all computations with the perturbative result, in this paper we shall therefore freely interchange the normal bundle
$\mathcal{N}_{S/B}$ of $S$ in the 3-fold base with the canonical bundle $K_{S}$.

As a brief aside, we caution that in some cases, a perturbative check may not be available.  For example, in the 7-brane configurations studied
in \cite{Beasley:2008dc, Beasley:2008kw}, the discriminant locus of the elliptic fibration enhances to an $E$-type
singularity at points of the del Pezzo surface wrapped by the GUT model 7-brane.  In this case,
the $SL(2,\mathbb{Z})$ action on the resultant bound state of $(p,q)$ 7-branes of different type is less straightforward.  It would be of
interest to extend the analysis of this paper to this more general case.  Nevertheless, in all of the cases we consider here,
a perturbative analysis of D3-instantons is applicable.

But even in those cases where a perturbative analysis is available, there appears to be some degree of confusion in the literature as to when
D3-instantons can contribute to the superpotential of the low energy theory.  The issue centers around the fact that
quantization of the $3$-$3$ open string sector would at first appear to be independent of the details
of the background 7-brane. In fact, the behavior of the Calabi-Yau 4-fold in the vicinity of a 7-brane
immediately shows that there is an important subtlety in this analysis.  To this end, in subsection \ref{subsec:Witten}, we
provide a heuristic argument which shows that the M-theory dual description of candidate D3-instantons satisfy Witten's arithmetic genus
constraint \cite{Witten:1996bn}.  This supports the conclusions of \cite{Akerblom:2006hx,Petersson:2007sc} that such instantons can contribute when the net flux on the 4-cycle is trivial and further suggests that they can continue to contribute even when this flux is nontrivial.  Along these lines, we present a slight generalization of the discussion of \cite{Akerblom:2006hx,Petersson:2007sc} in subsection \ref{subsec:KK33} which shows in quite concrete terms how
superpotential terms can arise in this case.
Appendix B contains additional details on the zero mode analysis of candidate
instanton configurations in the presence of appropriate internal fluxes.  Finally, in subsection \ref{6dINTERP} we briefly describe
D3-instanton contributions in intersecting 7-brane configurations with matter localized on a Riemann surface.  By appealing to
an interpretation in terms of higher dimensional gauge theory instantons, we explain on general grounds why we expect
D3-instantons to generate superpotential terms involving the fields localized on such matter curves.

\subsection{Witten's Arithmetic Genus Constraint}\label{subsec:Witten}

In \cite{Witten:1996bn}, Witten analyzed the contribution to the superpotential from Euclidean M5-branes in M-theory compactified on a Calabi-Yau 4-fold $X_4$.  In the absence of background fluxes, a necessary condition for an M5-brane to contribute is that the divisor $D$ wrapped by the M5-brane must satisfy
\begin{equation}
h^0(D,{\cal{O}}_D)-h^1(D,{\cal{O}}_D)+h^2(D,{\cal{O}}_D)-h^3(D,{\cal{O}}_D)=\chi(D,{\cal{O}}_D)=1\,,
\label{wittcond}\end{equation}
so that the arithmetic genus of $D$ vanishes.\footnote{The arithmetic genus $p_{a}(M)$ of an $n$-dimensional algebraic variety $M$ is defined as
$p_{a}(M) = (-1)^{n}(\chi(M,\mathcal{O}_{M})-1)$.} This constraint follows from a study of the $U(1)$ structure group of the normal bundle to $D$ within $X_4$.  Indeed, this $U(1)$ corresponds to the usual $R$-symmetry of the theory with respect to which the superpotential integration measure $d^2\theta$ has charge +1.  As shown in \cite{Witten:1996bn}, the classical instanton action $e^{-S_{\text{inst}}}$ of the M5-brane also carries a charge under this symmetry which is nothing other than $-\chi(D,{\cal{O}}_D)$.  As such, the condition \eqref{wittcond} is simply the requirement that the F-term $d^2\theta\,e^{-S_{\text{inst}}}$ be invariant under this $U(1)$ symmetry.

When $X_4$ is elliptically fibered, there exists an adiabatic limit which connects this compactification to F-theory compactified on the same $X_4$.  The corresponding instanton contributes to the superpotential of the four-dimensional theory when $D$ is a vertical divisor, so that two of its dimensions wrap directions in the elliptic fiber.  Now, although we are interested in configurations where the 3-brane has a nontrivial internal flux, the above condition should remain unchanged because the $3$-$3$ zero mode content, and hence the $U(1)$ charge of $e^{-S_{\text{inst}}}$, is insensitive to the choice of flux on the 3-brane.

In this paper we study configurations where 3-branes and 7-branes wrap the same K\"ahler surface $S$.  In all cases we consider, the presence of the background 7-branes plays a crucial role in the analysis.  To motivate this from geometry, suppose that in the vicinity of a divisor $S$ in the 3-fold base, $X_{4}$ is given by the product of a Calabi-Yau 3-fold and a smooth elliptic curve which we denote by $\mathbb{E}$.  In this case, the divisor $D$ factorizes as the product of $S$ and $\mathbb{E}$. The holomorphic Euler characteristic of $D$ now follows from the K\"unneth formula:
\begin{equation}
\chi(D,\mathcal{O}_D)=\chi(S,\mathcal{O}_{S})\cdot\chi(\mathbb{E},\mathcal{O}_\mathbb{E})=0\,.
\end{equation}
Thus, the D3-instanton cannot contribute to the 4d superpotential.

On the other hand, the divisor $D$ behaves quite differently when a 7-brane wraps $S$, because the elliptic curve is now singular along $S$.  To emphasize this point, we shall refer to the divisor of the Calabi-Yau 4-fold as $D_{\text{sing}}$.  For example, when a single 7-brane wraps $S$, the elliptic curve degenerates to an $I_1$ singularity which can be viewed as a $\mathbb{P}^1$ where the North and South poles touch.  At a somewhat naive level, we can view the divisor $D$ as the product of $S$ and a smooth $\mathbb{P}^{1}$, so that the holomorphic Euler character is given by:
\begin{equation}
\chi(D_{\text{sing}},{\cal{O}}_{D_{\text{sing}}})=\chi(S,{\cal{O}}_{S})\cdot\chi\left(\mathbb{P}^1,{\cal{O}}_{\mathbb{P}^1}\right)=\chi(S,{\cal{O}}_{S})\,.
\end{equation}
In other words, when a single 7-brane wraps a del Pezzo surface so that $\chi(S,{\cal{O}}_{S})=1$, the above argument would suggest that a Euclidean 3-brane could contribute to the 4d superpotential.

This type of argument can be made more precise in certain cases.  For example, as shown in \cite{Katz:1996th}, starting from the 7-brane theory with ADE gauge group defined by an appropriate ADE degeneration of the elliptic curve of $X_4$, in the dual M-theory picture, M5-instantons wrapping $\mathbb{P}^{1}$'s in the singular elliptic curve were summed and found to reproduce the expected dual description of the Veneziano-Yankielowicz \cite{Veneziano:1982ah} glueball superpotential.  In Section \ref{sec:puresun}, we shall compute additional contributions from the 1-instanton sector of the theory.

\subsection{3-3 Zero Modes Coupling to KK Modes}\label{subsec:KK33}

In the previous subsection we argued on geometric grounds that 3-branes wrapping the
same K\"ahler surface as a 7-brane can potentially contribute to the 4d superpotential. In
this subsection we directly study the zero mode content of such 3-branes in the vicinity
of a 7-brane and in particular explain the conditions under which such D3-instantons can contribute to the
4d superpotential.  For concreteness, we shall describe such instantons
in the context of type IIB string theory compactified on a Calabi-Yau 3-fold.

The counting of zero modes and hence the nature of superpotential couplings that can be generated by D3-instantons depends heavily on any worldvolume fluxes present on either the D3 or the D7's.  In this paper, it will be important to consider examples in which such fluxes are nontrivial as this will allow chirality in the zero mode spectrum and hence in the superpotential couplings themselves.  First, however, we shall review the zero mode counting when the fluxes on all branes are trivial.  This situation is quite well-studied in several examples and we summarize the implications for our system below.

\subsubsection{Zero Modes for Trivial Fluxes}

Let us begin with the 3-3 open strings that connect the D3-instanton to itself.  the 3-3 zero modes will always include four bosonic Goldstone modes $x^{\mu}$ associated to the spontaneous breaking of the 4-dimensional Lorentz symmetry, as well as four Goldstinos, $\theta_{\alpha},\mu_{\dot{\alpha}}$ with $\alpha,\dot{\alpha}=1,2$ associated to the four supercharges which are both preserved by the Calabi-Yau 3-fold and broken by the D3-instanton.  Because we will always consider instantons that wrap rigid 4-cycles, these universal zero modes are the only ones which are present.

The contribution of a D3-instanton to the 4d effective theory is given by integrating over all associated zero modes.
F-term contributions are given by integrating the appropriate tunneling amplitude over the measure factor $\int d^{4}x d^{2}\theta$.
This potentially leaves two additional universal $\mu_{\dot{\alpha}}$ fermionic zero modes to integrate over.  A non-zero
F-term contribution is possible if either this integration factor is absent, or is saturated in some way. Many techniques
are available for lifting these unwanted zero modes, including the introduction of orientifold planes and/or bulk fluxes
as described, for instance, in the review article \cite{Cvetic:2007sj} and references contained therein.

When D3-instantons are embedded within D7-branes, however, they
should be capturing the effects of 4-dimensional gauge theory
instantons, so we fully expect them to contribute in this case, as in \cite{Bershadsky:1996gx}.
This begs the question: What is taking care of the extra $\mu_{\dot{\alpha}}$ zero modes?
Heuristically, in this situation the D3-instantons are becoming
aware of the presence of the D7's, which are responsible for
reducing the supersymmetry of the background by half.  In
particular, the system of the Calabi-Yau plus D7-branes preserves only 4
supercharges of which only 2 are broken by the instanton.  This
means that the D3-instanton worldvolume should have only 2
Goldstinos and hence 2 fermi zero modes instead of 4.  One can
understand the \textquotedblleft lifting\textquotedblright of the
other two as a consequence of the D3 and D7's forming a bound state at threshold.

Let us make this more concrete.  As described in more detail in
Appendix \ref{app:pertcount}, the $3$-$3$ fermion zero modes of a
D3-instanton wrapping the 4-cycle $S$ are sections of{ \footnote{In
the following, we make extensive use of Serre duality and the fact
that $S$ is a del Pezzo surface.}}
\begin{equation}
\ba
\theta_\alpha \,:       &\qquad \left[S_+'\otimes H^2(S,K_S)\right] \cr
\mu_{\dot{\alpha}}\,:& \qquad \left[S_-'\otimes H^0(S,{\cal{O}})\right]\,,
\ea
\end{equation}
where $S_{\pm}'$ are the positive/negative chirality spin bundles of
$\mathbb{R}^{4}$ and $K_S$ is the canonical bundle of $S$. Because
$h^0(S,{\cal{O}})=h^2(S,K_S)=1$, this yields the usual
universal zero modes, namely a chiral $SO(4)$ spinor
$\theta_{\alpha}$ and an anti-chiral $SO(4)$ spinor
$\mu_{\dot{\alpha}}$.

As expected, the $3$-$3$ zero mode content does not appear to depend on whether a D7-brane is present.
However, the D7-brane will still influence the dynamics of the D3-instanton through the 3-7 and
7-3 open strings which connect the D3-instanton to the D7-brane.  Unlike the $3$-$3$ sector, these modes
will be charged under the gauge groups on the D3- and D7-branes and hence will transform in
some bifundamental representation $R$ (coming from 7-3 strings) and
its conjugate (coming from 3-7 strings). Because there is no net $U(1)$ flux on $S$,
the fermionic zero modes in the
representation $R$ are sections of
\begin{equation}
f\,, \ \bar{f}\, :\qquad
H^2(S,K_S) \,.
\end{equation}
Again, $h^2(S,K_S)=1$ so there is exactly one such zero mode
that we shall denote by $f$.  Similarly, we also get a single mode
$\bar{f}$ transforming in the conjugate representation $\bar{R}$.
Note that both of these modes are spacetime scalars, reflecting the
fact that fermions originate from NN and DD directions.  A similar analysis
establishes that we also obtain bosonic zero modes transforming in
the representations $R$ and $\bar{R}$.  These bosonic modes transform as sections of
\begin{equation}
b_{\dot\alpha}\,, \ \bar{b}_{\dot\alpha}\,:\qquad
S_-'\otimes H^0(S,{\cal{O}})\,.
\end{equation}
Note that because bosons come from ND directions, these are
spacetime spinors whose chirality is fixed by the GSO projection.

Because $h^0(S,\mathcal{O})=1$, we get exactly
one anti-chiral spinor worth of bosonic zero modes in the
representation $R$.  We shall denote this mode by $b_{\dot{\alpha}}$
and the corresponding mode which transforms in the
representation $\bar{R}$ by ${\bar b}_{\dot \alpha}$.  We summarize the
various zero modes (for trivial net $U(1)$ flux) which are present in
the following table

\begin{center}
\begin{tabular}{c|c|c|c}
Mode & Origin & Fermion/Boson & Gauge Rep \\ \hline
$x^{\mu}$ & 3-3 & Boson & $\mathbf{1}$\\
$\theta_{\alpha}$ & 3-3 & Fermion & $\mathbf{1}$ \\
$\mu_{\dot{\alpha}}$ & 3-3 & Fermion & $\mathbf{1}$ \\
$f$ & 7-3 & Fermion & $R$ \\
$\bar{f}$ & 3-7 & Fermion & $\bar{R}$ \\
$b_{\dot{\alpha}}$ & 7-3 & Boson & $R$ \\
$\bar{b}_{\dot \alpha}$ & 3-7 & Boson & $\bar R$
\end{tabular}\end{center}

Let us now consider how these modes couple to one another.
A crucial point is that, because of the spacetime chirality of the bosonic 3-7 and 7-3 zero modes,
they can couple only to the anti-chiral $3$-$3$ zero modes $\mu_{\dot{\alpha}}$
\begin{equation}
\mu_{\dot{\alpha}}\Bigl(b^{\dot{\alpha}}\bar{f} + {\bar
b}^{\dot \alpha}f\Bigr) \,,
\label{barthetaints}\end{equation}
and
\emph{not} the chiral ones $\theta_{\alpha}$.{\footnote{Note that there are additional terms in the full action, which can be found for example in \cite{Argurio:2007vqa}, but we shall focus our attention only on \eqref{barthetaints} because these are the only ones relevant for the discussion that follows.}}

As is evident from the interaction term of line \eqref{barthetaints},
the $\mu_{\dot{\alpha}}$ zero modes can combine with $\bar{f}$ or $f$ and lift from the massless
spectrum by turning on suitable expectation values for the bosonic
modes $b_{\dot{\alpha}}$ or ${\bar b}_{\dot \alpha}$. This corresponds to nothing
more than moving onto the Higgs branch of the theory. Indeed, just as in
the perhaps more familiar situation of the D0-D4 system, this describes the
formation of a D3-D7 bound state. On the other hand, when these
bosonic expectation values vanish, the D3 and D7 instead form a
marginal bound state at threshold.

Nevertheless, the D3-instanton can continue to contribute to the
superpotential even in this marginal situation. The reason for this is that
the $\mu_{\dot{\alpha}}$ integral can be saturated by bringing down
the interaction terms in \eqref{barthetaints}.  As shown first in
\cite{Billo:2002hm} and further reviewed in the more recent
literature \cite{Akerblom:2006hx}, this simply enforces the
well-known fermionic ADHM constraints for the corresponding gauge
theory instanton.{\footnote{We thank T. Weigand for explaining this
point to us.}}

\subsubsection{Zero Modes for Non-Zero Fluxes}

In the previous subsection we discussed possible instanton configuration in
the special case where the flux passing through the D3-brane and D7-brane
are both trivial. More generally, however, it is possible to consider
configurations where a flux threads these brane configurations. Indeed,
given a fixed flux configuration through the D7-brane, the net contribution
from Euclidean D3-branes about this background is given by performing a sum
over all possible fluxes for the Euclidean D3-brane. Of the candidate fluxes
through the Euclidean D3-brane, many will induce a mass term for the
fields $x^{\mu }$ of the worldvolume theory which describe the position of the purported
instanton in $%
\mathbb{R}
^{4}$. Said differently, the presence of a flux threading the Euclidean
D3-brane can remove candidate instanton configurations because the presence of flux can generate a large mass for 
the would be Goldstone modes. As we now explain, in order for a Euclidean D3-brane to contribute
as an instanton, the flux threading this D3-brane must lift to a trivial
class in the threefold base.\footnote{After the first version of our paper appeared, the analysis of 
\cite{Cvetic:2009mt} appeared which analyzed the explicit mechanism for zero mode saturation proposed in the present paper. This work prompted us 
to return to the present class of instanton contributions, and helped us to clarify the crucial 
role that a ``trivializable'' flux plays in the consistent formulation of possible instanton contributions.} 
The mechanism is related to a similar phenomenon present for D7-branes threaded by flux, which we now briefly review.

For a D7-brane wrapping a complex surface $S$, activating a flux in the
worldvolume of the D7-brane can induce a mass term for the $U(1)$ gauge
boson in four-dimensions. This is due to a coupling between the
4-form potential $C_{4}$ which propagates in ten-dimensions, and the field
strength, $F$, of the D7-brane:%
\begin{equation}
\int_{\mathbb{R}^{4}\times S}\,C_{4}\wedge F\wedge F\text{.}
\end{equation}%
Including also the kinetic term for $C_{4}$ and focusing on that part of $%
C_{4}$ that takes the form $C_{2}\wedge \omega _{2}$ for $\omega _{2}$
(resp. $C_{2}$) a 2-form on $B$ (resp. $\mathbb{R}^{4}$), we obtain the
following contribution to the 4-dimensional effective action
\begin{equation}
\int_{\mathbb{R}^{4}}\,\left\{ |dC_{2}|^{2}+dC_{2}\wedge A\,\left[
\int_{S}\omega _{2}\wedge F_{S}\right] \right\}
\end{equation}%
where $A$ is the 4-dimensional gauge boson and $F_{S}$ the internal field
strength. Integrating out $C_{2}$ leads to a mass for the gauge boson of the form:%
\begin{equation}
M_{U(1)}\propto \int_{S}\omega _{2}\wedge F_{S}\text{.}  \label{MASS}
\end{equation}%
As shown in \cite{Beasley:2008kw,Donagi:2008kj}, it is nevertheless possible
to retain a massless gauge boson in the four-dimensional effective theory
provided the right-hand side of equation (\ref{MASS}) vanishes. This amounts
to a topological condition on how the class of $F_{S}$ lifts to $B$. Indeed,
since $\omega _{2}$ descends from a harmonic form on $B$, whereas $F_{S}$ is
defined solely on $S$, this condition can be arranged provided $F_{S}$ lifts
to a trivial class in $B$. This point is especially crucial in an analysis
of GUT\ group breaking by a $U(1)_{Y}$ flux, and is similar in spirit to the
mechanism found in \cite{Buican:2006sn}. We shall refer to fluxes which lift
to trivial classes in $B$ as \textquotedblleft
trivializable\textquotedblright .

Next consider a Euclidean D3-brane with non-zero flux through its
worldvolume. Here, the presence of a non-zero flux can sometimes induce a
mass for the fields $x^{\mu }$ describing the position of the Euclidean
D3-brane in $%
\mathbb{R}
^{4}$. When this occurs, there is no \textquotedblleft instanton collective
coordinate\textquotedblright\ corresponding to a Goldstone mode, and thus no
instanton configuration to speak of. A necessary condition to have a
non-trivial $x^{\mu }$ collective coordinate is that the flux must not
induce a mass for these fields of the Euclidean D3-brane worldvolume theory.
To make contact with the discussion of D7-branes just given, consider a
Euclidean theory with $%
\mathbb{R}
^{4}$ replaced by a four torus, $T^{4}$. In this case, T-dualizing all four
directions of the D7-brane with internal flux corresponds to a D3-brane with
non-trivial internal flux, where $x^{\mu }$ is the T-dual of the
4-dimensional gauge field of the D7-brane theory. Since the relevant part of
$C_{4}$ has two legs in $T^{4}$ and two in $S$, it follows that there is a
T-dual 4-form potential, $\widehat{C}_{4}$ which couples to $x^{\mu }$
through the term:%
\begin{equation}
\int_{\mathbb{R}^{3,1}}\,\left\{ |d\widehat{C}_{2}|^{2}+d\widehat{C}%
_{2}\wedge x\,\left[ \int_{S}\omega _{2}\wedge F_{S}\right] \right\} ,
\end{equation}%
where in the above, we have written the part of $\widehat{C}_{4}$ which
takes the form $\widehat{C}_{2}\wedge \omega _{2}$ for $\omega _{2}$ (resp. $%
\widehat{C}_{2}$) a 2-form on $B$ (resp. $T^{4}$). Integrating out $\widehat{%
C}_{2}$ then induces a mass term for the fields $x^{\mu }$. Since this
coupling clearly persists when $T^{4}$ is replaced by $%
\mathbb{R}
^{4}$, we see that the same mechanism which lifts a massless gauge boson
from the D7-brane theory can also induce mass terms for the $x^{\mu }$
fields of the Euclidean D3-brane worldvolume theory. In order to retain a
massless mode, the resolution is also the same as in the case of the
D7-brane theory; the flux threading the D3-brane must be trivializable in
the threefold base $B$.

Hence, in studying possible instanton contributions from Euclidean D3-branes
with non-trivial flux, only flux through the D3-brane with ``trivializable''
flux can contribute. It is important to note, however, that the flux through
the D7-brane need not be trivial. Indeed, the only effect that this will
have is to remove the corresponding $U(1)$ gauge boson from the low energy
spectrum.

The zero mode spectrum for the instanton configuration is controlled by the
flux through the D3- and D7-branes of the configuration. Note that even if the flux through the D3-brane is
``trivializable'', it is still possible to generate a chiral zero mode spectrum
because the D7-brane flux need not be ``trivializable''. The most important
implication of such a flux is that it changes the counting of 3-7 and 7-3
zero modes. In particular, the modes $f,\overline{f},b^{\dot{\alpha}},%
\overline{b}^{\dot{\alpha}}$ that were used to saturate the $\mu _{\dot{%
\alpha}}$ integration in \eqref{barthetaints} are lifted. This completely
lifts the Higgs branch, thereby preventing the formation of an honest D3-D7
bound state. Nevertheless, the marginal bound state from which we were able
to derive a nonzero superpotential contribution above still remains. One can
therefore ask whether we still expect a contribution from this marginal
situation even when the honest Higgs branch ceases to exist.

Even in this case, however, the correct way to deal with the extra $\mu_{%
\dot{\alpha}}$ 3-3 zero modes is to proceed precisely as before. In
particular, we note that even though there are no 3-7 or 7-3 zero modes for $%
\mu_{\dot{\alpha}}$ to couple to, we should recall that \eqref{barthetaints}
represents in fact a Kaluza-Klein reduction of parent interactions in the
full 8-dimensional theory on the D7-brane worldvolume. As such, we expect
that $\mu_{\dot{\alpha}}$ has similar couplings to analogous modes in the KK
tower. To write the full action for these modes, however, it is important to
recall how they couple to one another. The full set of 3-7 and 7-3 modes
descend from various $(p,q)$-forms on the internal space as described in
Appendix \ref{app:ZeroModes}. We summarize the relevant results in the
following table
\begin{equation}
\begin{array}{c|c|c|c|c}
\text{Field} & \text{Origin} & \text{Fermion/Boson} & (p,q) & \text{%
Representation} \\ \hline
\mu_{\dot{\alpha}} & 3-3 & F & (0,0) & 1 \\
b^{\dot{\alpha}}_{KK} & 3-7 & B & (0,0) & R \\
\overline{b}^{\dot{\alpha}}_{KK} & 7-3 & B & (0,0) & \overline{R} \\
f^{(2,0)}_{KK} & 3-7 & F & (2,0) & R \\
f^{(0,1)}_{KK} & 3-7 & F & (0,1) & R \\
f^{(2,2)}_{KK} & 3-7 & F & (2,2) & R \\
\overline{f}^{(0,2)}_{KK} & 7-3 & F & (0,2) & \overline{R} \\
\overline{f}^{(1,0)}_{KK} & 7-3 & F & (1,0) & \overline{R} \\
\overline{f}^{(2,2)}_{KK} & 7-3 & F & (2,2) & \overline{R}
\end{array}%
\end{equation}
The modes whose coupling to $\mu_{\dot{\alpha}}$ can saturate the $d^2\mu$
integration are precisely the KK partners of the ones in \eqref{barthetaints}%
, namely $b^{\dot{\alpha}}_{KK},\overline{b}^{\dot{\alpha}%
}_{KK},f^{(2,2)}_{KK},\overline{f}^{(2,2)}_{KK}$. Further, the only sensible
mass terms that can involve these modes are $\overline{b}_{KK}b_{KK}$ and $%
\overline{f}^{(2,2)}_{KK}f^{(2,2)}_{KK}${\footnote{%
Of course an $\epsilon$ tensor must be used to construct these.}}. Combining
all of these, we obtain the following{\footnote{%
As described below, this action can be understood as the only one that is
invariant under the topologically twisted $U(1)_{top}$.}}
\begin{equation}
\mu_{\dot{\alpha}}\left(b^{\dot{\alpha}}_{KK}\overline{f}^{(2,2)}_{KK}+%
\overline{b}^{\dot{\alpha}}_{KK}f_{KK}^{(2,2)}\right)+M_{b,KK}\overline{b}%
_{KK}b_{KK}+M_{f,KK}\overline{f}^{(2,2)}_{KK}f^{(2,2)}_{KK}  \label{KKaction}
\end{equation}
which is more than sufficient for the purpose of saturating the $d^2\mu$
integral. Although this slightly modifies the measure factor for the KK
modes, the net effect of this contribution can typically be absorbed into an
overall multiplicative factor which is for the most part unimportant.

Taken together, this reasoning leads us to conclude that, whenever a
D3-instanton is embedded within a 7-brane, it forms a bound state at
threshold and can in general yield a nonzero superpotential contribution to
the worldvolume theory regardless of what supersymmetric $U(1)$ flux is
present on the 7-brane. To obtain the full instanton-generated
superpotential, though, it is also necessary to sum over the contributions
of all ``trivializable'' $U(1)$ fluxes that can be introduced on the instanton.

\subsubsection{Relation to the Results of \cite{Cvetic:2009mt}}

Before closing this subsection, we now comment on the relation of our results to those of \cite{Cvetic:2009mt}, where a similar set of instanton contributions were studied in the context of type IIA intersecting brane models.  More specifically, \cite{Cvetic:2009mt} considers an example in which the only available means to deal with the $\mu_{\dot{\alpha}}$ zero modes is via a mechanism similar to the one described here.  That is, the only hope of a nontrivial contribution to the superpotential is to saturate the $\mu_{\dot{\alpha}}$ integrations through couplings involving massive modes that connect the D-brane instanton to a space-filling D-brane.  In their specific example, a null result was obtained because the effective action for the relevant modes was not of the right form to allow the $\mu_{\dot{\alpha}}$ integrations to be saturated.  Here we would like to point out that this is due to the presence of additional $U(1)$ R-symmetries in their example which are not present in the class of instanton configurations considered in the present paper.  In this regard, an especially crucial role is played by the fact that in the present context, ${\cal{N}}=1$ supersymmetry dictates a unique partial twist of the seven-brane gauge theory.


In order to make a connection to the computation of \cite{Cvetic:2009mt},
let us now see directly how the twisting enters into our previous arguments
concerning KK modes. In general, if there is no twist, the 3-3 zero modes
descend from chiral spinors of $SO(9,1)$ that we can further decompose under
$SO(9,1)\rightarrow SO(3,1)\times U(2)\times U(1)_{R}$, where $U(2)$ is the
local isometry of $S$ and $U(1)_{R}$ is that of the transverse direction.
The anti-chiral 3-3 modes, $\mu _{\dot{\alpha}}$, transform under $%
U(2)\times U(1)_{R}$ as
\begin{equation}
\mu _{\dot{\alpha}}:\quad \left( \mathbf{1}_{+1}\oplus \mathbf{1}_{-1},+%
\frac{1}{2}\right) \oplus \left( \mathbf{2}_{0},-\frac{1}{2}\right)
\end{equation}%
If there are any $\mu _{\dot{\alpha}}$ zero modes, our goal is to saturate
their integration through couplings of the form
\begin{equation}
\mu _{\dot{\alpha}}b^{\dot{\alpha}}\overline{f} + \mu_{\dot{\alpha}}\overline{b}^{\dot{\alpha}}f  \label{mudotcoupling}
\end{equation}%
where $b^{\dot{\alpha}}$ $(\overline{b}^{\dot{\alpha}})$ and $f$ ($\overline{f}$) bosonic and fermionic modes arising from
3-7 (7-3) strings. The integration over $\mu _{\dot{\alpha}}$ will in
general bring down two factors of fermi modes, say $f_{1}f_{2}$. If $f_{1}$
and $f_{2}$ are zero modes themselves, then they are simply absorbed by the $%
df_{1}$ and $df_{2}$ integrations with no problem. On the other hand, if $%
f_{1}$ and $f_{2}$ are massive modes, then it is necessary that they become
massive in part due to a quadratic coupling of the form $f_{1}f_{2}$ in the
action{\footnote{%
If $f_{1}$ instead becomes massive by coupling to a different mode $\tilde{f}%
_{1}$ and $f_{2}$ by coupling to $\tilde{f}_{2}$ then it is impossible to
saturate the integrations over all of $f_{1}$, $f_{2}$, $\tilde{f}_{1}$, and
$\tilde{f}_{2}$ in the presence of the $f_{1}f_{2}$ insertion.}}. This means
that the object $f_{1}f_{2}$ must be invariant under $U(2)\times U(1)_{R}$.
Note, however, that the bosonic 3-7 and 7-3 modes descend from antichiral
spinors under $SO(3,1)$ and hence are $U(2)\times U(1)_{R}$ singlets. This
means that in order for $f_{1}$ and $f_{2}$ to couple to $\mu _{\dot{\alpha}}
$ as in \eqref{mudotcoupling}, they must transform in the same way under $%
U(2)\times U(1)_{R}$, namely the representation conjugate to the particular
mode of $\mu _{\dot{\alpha}}$ to which they couple. The $U(2)$
representation of the object $f_{1}f_{2}$ is therefore either $\mathbf{1}%
\otimes \mathbf{1}$ or $\mathbf{2}\otimes \mathbf{2}$ so a $U(2)$ singlet of
this form always exists. However, all modes $\mu _{\dot{\alpha}}$ carry
nonzero $U(1)_{R}$ charge, which means that the object $f_{1}f_{2}$ is also
guaranteed to carry a nonzero $U(1)_{R}$ charge. Such a quadratic coupling
is therefore prohibited and suitable interaction terms \eqref{mudotcoupling}
of the right structure to allow a nonzero superpotential coupling cannot
occur.

The mechanism that prevents saturation of the $\mu _{\dot{\alpha}}$ modes in
this case is in fact precisely the one that arose in the computation of \cite%
{Cvetic:2009mt}. To see this, let us describe the precise structure of
quadratic and cubic couplings in this case. Fermi modes arise from NN and DD
directions so the 3-7 fermi modes, which transform in the gauge
representation $R$, comprise a chiral spinor under the $SO(6)$ associated to
the internal directions of $B$. Similarly, the 7-3 fermi modes transform in
the gauge representation $\overline{R}$ and comprise an antichiral spinor
under $SO(6)$. Under the decomposition $SO(6)\rightarrow U(2)\times U(1)_{R}$%
, the 3-7 and 7-3 fermi spectrum can be written as
\begin{equation}
3-7\text{ Fermi}:\quad \left( \mathbf{2}_{0}\oplus \mathbf{1}_{+1}\oplus
\mathbf{1}_{-1},+\frac{1}{2}\right) \qquad 7-3\text{ Fermi}:\quad \left(
\mathbf{2}_{0}\oplus \mathbf{1}_{+1}\oplus \mathbf{1}_{-1},-\frac{1}{2}%
\right)
\end{equation}%
Similarly, the 3-7 and 7-3 bosonic modes arise from ND directions, namely
from $\mathbb{R}^{3,1}$. They correspond to antichiral spinors under $SO(3,1)
$ in the representations $R$ and $\overline{R}$, respectively. The $%
U(2)\times U(1)_{R}$ reps are therefore trivial
\begin{equation}
3-7\text{ Bose}:\quad \mathbf{1}_{0}\qquad 7-3\text{ Bose}:\quad \mathbf{1}%
_{0}
\end{equation}%
We see that each type of $\mu _{\dot{\alpha}}$ mode has a unique 3-7 or 7-3
Fermi mode to which it can couple. For instance, the $\mu _{\dot{\alpha}}$
mode $(\mathbf{1}_{+1},+1/2)$ can couple to the 7-3 fermi mode $(\mathbf{1}%
_{-1},-1/2)$. This fermi mode becomes massive by pairing up with the 3-7
mode $(\mathbf{1},+1,+1/2)$ which, in turn, cannot couple to any modes of $%
\mu _{\dot{\alpha}}$. We therefore get the following structure of couplings
\begin{equation}
\mu _{\dot{\alpha}}b^{\dot{\alpha}}\overline{f}+Mf\overline{f}
\end{equation}%
where $f$ is forbidden to couple with $\mu _{\dot{\alpha}}$. This
is precisely the structure found in the explicit intersecting brane example
of \cite{Cvetic:2009mt}, which is therefore an effective model when
additional $U(1)$ R-symmetries are present.

In the class of examples considered here, however, there is a unique partial
twisting available, so only a particular combination of $U(1)_{R}$ and the $%
U(1)_{J}$ subgroup of $U(2)$ survives as a symmetry of the system. This is
the topologically twisted $U(1)$, denoted $U(1)_{top}$, which is related to $%
U(1)_{J}$ and $U(1)_{R}$ by $J_{top}=J+2R$. The 3-3 modes $\mu _{\dot{\alpha}%
}$ are antichiral spinors of $SO(3,1)$ which transform under $SU(2)\times
U(1)_{top}$ as
\begin{equation}
\mu _{\dot{\alpha}}:\quad \left( \mathbf{2}_{-1}\oplus \mathbf{1}_{+2}\oplus
\mathbf{1}_{0}\right)
\end{equation}%
Moreover, in our example, the $\mu _{\dot{\alpha}}$ zero modes on $S$
correspond to the $\mathbf{1}_{0}$ representation. Because it is uncharged
under $U(1)_{top}$, various fermi modes $f_{1},f_{2},\ldots $ to which it
can couple are also uncharged and hence can generically exhibit quadratic
couplings in the action. More specifically, the 3-7 and 7-3 Fermi modes
transform under $SU(2)\times U(1)_{top}$ as
\begin{equation}
3-7\text{ Fermi}:\quad \left( \mathbf{2}_{+1}\oplus \mathbf{1}_{-2}\oplus
\mathbf{1}_{0}\right) \qquad 7-3\text{ Fermi}:\quad \left( \mathbf{2}%
_{-1}\oplus \mathbf{1}_{+2}\oplus \mathbf{1}_{0}\right)
\end{equation}%
The $\mathbf{1}_{0}$ mode of $\mu _{\dot{\alpha}}$ can only couple to the $%
\mathbf{1}_{0}$ 3-7 and 7-3 fermi modes, which correspond to $f_{KK}^{(2,2)}$
and $\overline{f}_{KK}^{(2,2)}$ in \eqref{KKaction}. Furthermore, the only
invariant mass term involving $f_{KK}^{(2,2)}$ and $\overline{f}_{KK}^{(2,2)}
$ is precisely the quadratic combination $f_{KK}^{(2,2)}\overline{f}%
_{KK}^{(2,2)}$. In that sense, the action \eqref{KKaction} contains all
terms involving $\mu _{\dot{\alpha}}$, $f_{KK}^{(2,2)}$, and $\overline{f}%
_{KK}^{(2,2)}$ that are invariant under $U(1)_{top}$. From symmetries alone,
then, we see that the action is forced to have the right structure to allow
saturation of the $\mu _{\dot{\alpha}}$ zero modes.

\subsection{Intersecting 7-Branes and Higher Dimensional Instantons}

\label{6dINTERP}

In the previous subsection we showed that the theory defined by an isolated
7-brane wrapping a rigid K\"ahler surface contains the expected contribution
from 4d gauge theory instantons, and that these contributions can be
identified with D3-instantons wrapping the same surface as the 7-brane. More
generally, we can also consider configurations where distinct 7-branes
intersect over a Riemann surface $\Sigma$. This case is analyzed in detail
in Appendix B and forms the basis for the Polonyi-like model of Section \ref%
{sec:polonyi}. We now explain from the perspective of the 6d theory on $%
\mathbb{R}^{4}\times \Sigma$ why we generically expect instanton corrections
to the 4d superpotential in this case as well.

For simplicity, consider a configuration of two intersecting 7-branes with
gauge group $U(1)\times U(1)$ and bifundamental matter localized at the
intersection. As reviewed in Appendix B, the 4d chiral matter content
depends on the background flux on the 7-branes. If there is an excess of
chiral matter of one type, the 4d gauge theory will be anomalous, and the
corresponding gauge bosons will lift from the low energy spectrum via the
Green-Schwarz mechanism. The low energy theory will then contain an
anomalous global symmetry which will generically be violated by instanton
effects.

While this might suggest that a non-anomalous 4d gauge theory will not
receive D3-instanton corrections, in a higher dimensional sense, this theory
is always anomalous because the 6d theory defined by reduction of the 8d
gauge fields of the 7-branes to $\mathbb{R}^{4}\times \Sigma $ couples to a
single 6d chiral field. In particular, any higher dimensional instanton will
detect this anomaly if $TrF^{3}$ is non-trivial on the 6d spacetime $\mathbb{%
R}^{4}\times \Sigma $. This is completely in accord with an interpretation
in terms of D3-instantons. Indeed, as explained at the beginning of this
Section, we may view this D3-instanton as a non-trivial contribution to $%
TrF^{2}$ in $\mathbb{R}^{4}$, and the internal flux through the D3-brane
simply reflects the fact that a D1-brane can bind to this configuration. All
told, the Euclidean D1-D3 bound state is precisely what is described by the
non-trivial Chern characters $TrF^{2}$ and $TrF^{3}$ on $\mathbb{R}%
^{4}\times \Sigma $. This has the consequence that even if the 4d gauge
theory is non-anomalous, the presence of the 6d anomaly indicates that we
should still expect instanton contributions to the superpotential.

More generally speaking, the instantons of interest correspond to 
BPS\ field configuration defined by the configuration
of intersecting seven-branes. This point of view is especially helpful in
determining precisely which types of field configurations are compatible
with the presence of $\mathcal{N}=1$ supersymmetry in four dimensions. The
essential point is that in eight dimensions, a BPS field configuration must
satisfy the supersymmetric condition:%
\begin{equation}
\delta _{\alpha }\Psi _{\text{gaugino}}=0,
\end{equation}%
where $\Psi _{\text{gaugino}}$ denotes the gaugino of the eight-dimensional
gauge theory, and $\delta _{\alpha }$ denotes variation with respect to one
of the supercharges of the four-dimensional effective theory, with $\alpha
=1,2$ a spinor index in $\mathbb{R}^{4}$. In the context of the partially
twisted 7-brane theory, there are four real supercharges corresponding to $%
(0,0)$ forms on $S$. With respect to this unique twist, $\Psi _{\text{gaugino%
}}$ decomposes into a $(0,0)$, $(0,1)$ and $(2,0)$ form on $S$. As an
example, the variation of the $(0,0)$ form $\eta _{\beta }$ and its
conjugate $\overline{\eta }_{\dot{\beta}}$ are \cite{Beasley:2008dc}:%
\begin{eqnarray}
\delta _{\alpha }\eta _{\beta } &=&\left( \sigma ^{\mu \nu }\right) _{\alpha
\beta }F_{\mu \nu }-i\epsilon _{\alpha \beta }\mathcal{D} \\
\delta _{\alpha }\overline{\eta }_{\dot{\beta}} &=&0,
\end{eqnarray}%
where:%
\begin{equation}
\mathcal{D}=-\ast _{S}\left( J\wedge \left[ F_{S}-\frac{1}{2}\delta _{\Sigma
}\mu \left( \sigma ,\sigma \right) +\frac{1}{2}\delta _{\Sigma }\mu \left(
\sigma ^{c},\sigma ^{c}\right) \right] \right) ,  \label{DtermDterm}
\end{equation}%
where $F_{S}$ denotes the internal field strength, $\sigma $ and $\sigma ^{c}
$ denote fields on the matter curve $\Sigma $, and $\mu $ denotes a moment
map so that schematically $\mu \left( \sigma ,\sigma \right) \sim \left\vert
\sigma \right\vert ^{2}$ in the 4-dimensional effective theory. In the
context of an anomalous gauge theory where the number of $\sigma $ and $%
\sigma ^{c}$ zero modes differ, bulk axion-like modes can effectively play
the role of additional $\sigma $'s or $\sigma ^{c}$'s, so that there is a close analogue of 
equation (\ref{DtermDterm}) in this case as well.

From the perspective of the four dimensional theory on $%
\mathbb{R}
^{4}$, a Euclidean D3-brane wrapping $S$ can be viewed as a field
configuration $F_{\mu \nu }$ satisfying the usual Hermitian Yang-Mills
equations. In particular, this requires:%
\begin{equation}
\left( \sigma ^{\mu \nu }\right) _{\alpha \beta }F_{\mu \nu }=0.
\label{HYM4d}
\end{equation}%
Compatibility with $\delta _{\alpha }\eta _{\beta }=0$ then implies that:%
\begin{equation}
\mathcal{D}=0,  \label{Dcond}
\end{equation}%
which is nothing other than the usual D-term equation of motion, with $%
J\wedge F_{S}$ integrated over $S$ playing the role of an effective FI\
parameter. Interpreting the Euclidean D3-brane bound to the D7-brane in terms of a higher 
dimensional BPS configuration of the eight-dimensional partially twisted theory, the 
net flux $F_{S}$ is then given as a contribution from the flux threading 
the D3-brane, as well as a flux threading the D7-brane with:%
\begin{equation}
F_{S}=F^{(D3)}+F^{(D7)}\text{.}
\end{equation}

\subsubsection{Effective Field Theory Considerations}

When discussing potentially novel superpotential couplings induced by
stringy instantons, it is important to ensure that the results are
consistent with standard holomorphy arguments for contributions to the
superpotential of the 4d effective field theory. For example, holomorphy
considerations imply that deformations of the K\"ahler form cannot cause an
instanton to simply disappear; Indeed, if there exists a region in K\"ahler
moduli space where the superpotential vanishes, then holomorphy implies that
the superpotential coupling is simply not generated by instanton effects.
See for example \cite{GarciaEtxebarria:2007zv,GarciaEtxebarria:2008pi} for
applications of wall crossing phenomena to possible instanton generated
superpotential terms. In this context, it is sometimes convenient to phrase
the condition for the existence of an appropriate BPS\ configuration in $%
\mathcal{N}=2$ language as the condition that the phases of the central
charges for the D3-brane and D7-brane must align under K\"ahler deformations. 
Indeed, when the phases of the two central charges
become misaligned, the D3-brane will contribute four, rather two Goldstino
modes and a superpotential coupling would not be generated. As a brief 
aside, we note that there are various caveats to such
holomorphy arguments. For example this does not preclude the existence of
walls of marginal stability where a 1-instanton configuration could break up
into a 2-instanton configuration.

Similar considerations also apply in the present class of instanton configurations. 
In particular, equation (\ref{HYM4d}) requires that any
instanton configuration must satisfy equation (\ref{Dcond}). Said
differently, if there exists a K\"{a}hler deformation such that $\mathcal{D}%
\neq 0$, then the instanton equation of motion will be deformed, so that the
given field configuration cannot contribute to the superpotential. Since a
candidate bound state of a D3-brane and D7-brane can be viewed in terms of a
BPS\ field configuration of an eight-dimensional gauge theory, it is enough
to analyze whether there exist K\"{a}hler deformations which can violate the
condition $\mathcal{D}=0$.

Returning to equation (\ref{DtermDterm}), the condition $\mathcal{D}=0$
requires:

\begin{equation}
J\wedge \left[ F^{(D7)}-\frac{1}{2}\delta _{\Sigma }\mu \left( \sigma
,\sigma \right) +\frac{1}{2}\delta _{\Sigma }\mu \left( \sigma ^{c},\sigma
^{c}\right) \right] =0,
\end{equation}%
where $F^{(D7)}$ denotes the internal flux threading the D7-brane. Here, we
have used the fact that $J\wedge F^{(D3)}=0$ since $F^{(D3)}$ is
trivializable in the threefold base $B$. Note that even if $J \wedge F^{(D7)} \neq 0$, 
the vevs of the $\sigma$'s in the D-term equation will in general adjust so that the condition 
$\mathcal{D} = 0$ is retained.

These considerations will prove important in the two specific applications
presented in Sections 3 and 4. In section 3, we will be interested in field
configurations such that $J\wedge F^{(D7)}=0$, but where $F^{(D7)}$ is not
trivializable in the threefold base $B$. In those cases where $F^{(D7)}$ is
not ``trivializable'', there exists a direction in the K\"{a}hler cone $%
J^{\prime }$ such that $J^{\prime }\wedge F^{(D7)}\neq 0$. Thus, even small
perturbations of $J$ of the form $J+\varepsilon J^{\prime }$ for $%
\varepsilon $ small can produce a K\"ahler form which is not orthogonal to $F_S$. Even in
this more general case, however, the discussion above shows that there still exists a supersymmetric field
configuration such that the condition $\mathcal{D}=0$ is met. 
Thus, the BPS configuration still retains precisely two Goldstino modes
under the deformation of the K\"{a}hler potential, in accord with holomorphy
considerations. Finally, in section 4 we will consider another possible
instanton contribution where the flux threading the D7-brane is zero, and
that from the D3-brane is ``trivializable''. In this case, there are no 
bifundamentals, but also no K\"{a}hler deformation available to consider.


\section{A Polonyi-Like Model}
\label{sec:polonyi}

In this Section we present a simple realization of a Polonyi-like model based on intersecting 7-branes.  The Polonyi superpotential is generated by a D3-instanton embedded into one of the 7-branes, leading to an exponentially suppressed SUSY-breaking scale.  In particular, we show that the net contribution from all flux sectors of the 1-instanton theory is given by the sum of a term which only depends on the volume modulus, and another which is linear in $X$. After showing that this result is consistent with constraints from axion shifts, we study subleading corrections to this result due to multi-instanton effects.  In this Section we shall assume as in \cite{Beasley:2008dc, Beasley:2008kw} that this closed string mode has been stabilized by some high scale dynamics and we shall therefore treat it as a constant in the Polonyi-like model.  Nevertheless, the dynamics of this mode is quite interesting in its own right, and in Section \ref{sec:puresun}, we will study some consequences of this for pure 4d $\mathcal{N} = 1$ $SU(N)$ gauge theory engineered from a stack of 7-branes.

\subsection{General Setup}
\label{ourpolonyi}

\begin{figure}
\begin{center}
\epsfig{file=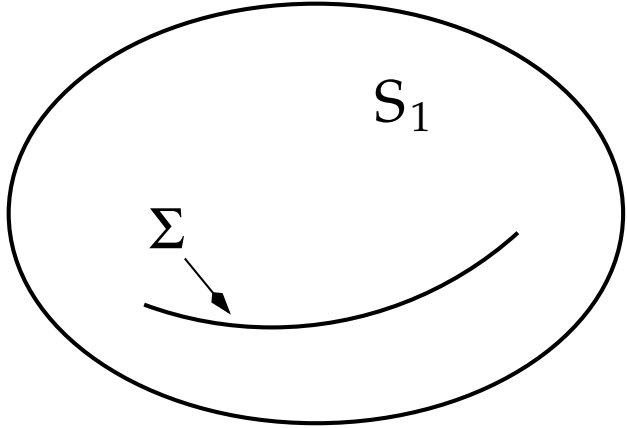,width=0.5\textwidth}
\caption{Basic Setup for Engineering Polonyi}
\label{polbranefig}
\end{center}
\end{figure}


We now present the basic setup for our Polonyi-like model.  The
model consists of two intersecting D7-branes wrapped
on 4-cycles $S_1$ and $S_2$ which intersect over a curve $\Sigma$.
To simplify the analysis to follow, we further assume that $S_2$
has much larger volume than $S_1$ so that it is enough to only consider D3-instantons wrapped on $S_1$. See
figure \ref{polbranefig} for a depiction of this intersecting brane configuration.

As briefly explained in the Introduction, D3-instanton contributions to surfaces of general
type are typically quite subtle because saturating all necessary zero modes requires the presence of a background flux.
In the spirit of \cite{Beasley:2008dc,Beasley:2008kw}, we shall therefore assume that the $S_a$'s are del Pezzo surfaces.  Even though
$S_{1}$ is the smaller of the two surfaces, we shall typically assume in our local model that the K\"ahler form $J$ on
$S_1=dP_M$  is fixed in the class
\be \label{kahler}
J =AH+\sum_{i=1}^M B_iE_i \,,\qquad A \gg 1\,, \quad B_{i}<0,\quad \vert B_{i}\vert \sim O(1), \quad i=1,\cdots,M  \,.
\ee
The 4d chiral matter content localized on $\Sigma$ is determined by a choice of background line bundles $V_{a}$ on the 7-branes wrapping the $S_{a}$'s.  A supersymmetric line bundle $L$ on $S_1$ must satisfy
\be
\int_{dP_M} c_1(L)\wedge J =0 \,.
\ee
For our present purposes, we do not need to specify explicitly the K\"ahler form on $S_2$ (which may be non-compact in the local model)
since in our analysis below we only use the restriction of $V_2$ to $\Sigma$.

The 4d zero mode content of the theory consists of a $U(1)$ vector multiplet for each 7-brane. In addition, as derived in \cite{Beasley:2008dc}, a particular choice of background fluxes on the 7-branes will determine the number of bifundamental matter fields charged under $U(1)_1\times U(1)_2$.  In a perturbative description, these modes correspond to open strings which begin on one 7-brane and end on the other 7-brane.\footnote{We note that in general, the ends of an open string will be twisted by $\pm \half K_{S}$ \cite{Minasian:1997mm}.  The effect of this in the topological theory is that the restriction of the bulk line bundle on each $S_{a}$ will receive an additional half integral shift by the restriction of the canonical class of each surface to $\Sigma$.  As in \cite{Beasley:2008dc, Beasley:2008kw}, this subtlety can be consistently neglected by an appropriate redefinition of the line bundle on the effectively non-compact 7-brane wrapping $S_{2}$.  Alternatively, we can always restrict to geometries where $K_{S_{1}}|_{\Sigma} = K_{S_{2}}|_{\Sigma}$, where this subtlety is absent.} We will see in a moment that with a suitable choice of bundles on $S_i,$ we can obtain a single chiral multiplet $X$ with charges $(+,-)$ under $U(1)_1\times U(1)_2$.  This will be the SUSY-breaking field of our Polonyi model.  With this in hand, we will set out in the rest of this Section to generate the Polonyi superpotential
\begin{equation}
W_P = F_X X\,,
\label{WPinst}\end{equation}
from a D3-instanton wrapping $S_1$.


\subsection{Generating Polonyi with D3-Instantons}\label{genpol}
We now proceed to the main aim of this Section which is to show that a D3-instanton wrapping $S_{1}$ can generate the Polonyi superpotential of equation \eqref{WPinst}. As described in Section \ref{sec:generalities}, we will allow the D3-instanton to house a nontrivial supersymmetric bundle $V_{\text{inst}}$ of the
$U(1)_{\text{inst}}$ gauge group on its worldvolume.  Indeed, to determine the net effect of D3-instantons we must effectively sum over all internal flux sectors of the worldvolume theory defined by the D3-instanton.  To determine the superpotential generated by the D3-instanton, we will make extensive use of the discussion in Appendix B, where we summarize how to count massless matter living on the intersection of D7-branes and various fermion zero modes living on the D3-instanton.


\subsubsection{Can a Superpotential be Generated?}

Our first task is to ensure that the instanton can contribute at all.  While we discussed this in generality in Section \ref{sec:generalities}, let us describe it in detail for the simple system at hand.  For this, we first count the $3$-$3$ fermi zero modes connecting the instanton to itself as{\footnote{Here, we use the fact that $H^0(S,{\cal{O}})$ and $H^2(S,K_S)$ are the only non-trivial cohomology classes for the trivial adjoint bundle on del Pezzo surfaces}}
\begin{equation}
\left[S_+'\otimes H^2(S_1,K_{S_1})\right]\oplus\left[S_-'\otimes H^0(S_1,{\cal{O}})\right] \,,
\end{equation}
where $S_+'$ ($S_-'$) denotes the spin bundle of positive (negative) chirality in $\mathbb{R}^{4}$.  Since $h^0(S_1,{\cal{O}})=h^2(S_1,K_{S_1})=1$ we obtain the usual universal zero modes consisting of a chiral spinor $\theta_{\alpha}$ and an anti-chiral spinor $\mu_{\dot{\alpha}}$.

As described in Section \ref{sec:generalities}, the integration over $\mu_{\dot{\alpha}}$ is saturated by coupling to bosonic and fermionic modes connecting the instanton to the D7-brane.  The necessary bosonic modes are sections of $S_-'\otimes\Omega^{0,0}\otimes V_1^{-1}\otimes V_{\text{inst}}$ and $S_-'\otimes \Omega^{0,0}\otimes V_1\otimes V_{\text{inst}}^{-1}$.  It is a consequence of the vanishing theorem in \cite{Beasley:2008dc} that there are no zero modes of the Laplacian acting on these sections, but there are nevertheless KK modes which we denote by $\bar{b}^{\dot{\alpha}}_{KK}$ and $b^{\dot{\alpha}}_{KK}$, respectively.  The necessary fermionic modes, on the other hand, are sections of $\Omega^{0,2}\otimes K_{S}\otimes V_1\otimes V_{\text{inst}}^{-1}$ and $\Omega^{0,2} \otimes K_{S}\otimes V_1^{-1}\otimes V_{\text{inst}}$.  Again, there are no zero modes of the Laplacian acting on these sections, but there are KK modes $f_{KK}$ and $\bar{f}_{KK}$.  The various KK modes couple to the zero modes $\mu_{\dot{\alpha}}$ through the interaction terms
\begin{equation}
\mu_{\dot{\alpha}}\left(\bar{b}^{\dot{\alpha}}_{KK}f_{KK}+b^{\dot{\alpha}}_{KK}\bar{f}_{KK}\right)+
M_{b,KK} {\bar b}_{KK}b_{KK}+M_{f,KK} {\bar f}_{KK}f_{KK} \,,
\label{P33couplings}
\end{equation}
which are sufficient to saturate the integral over the zero modes $\mu_{\dot{\alpha}}$.


\subsubsection{General Conditions on Bundles}

Now that we have established that D3-instantons can in principle yield a nonzero superpotential contribution, we proceed to determine a set of necessary conditions for fluxes through the D7-branes and D3-instanton to generate the Polonyi superpotential of equation \eqref{WPinst}.  In subsequent subsections we will show that this is in fact the only non-vanishing contribution from the D3-instanton. To set notation, we let $n_{pqr}$ denote the number of zero modes with charges $(p,q,r)$ under $U(1)_1\times U(1)_2\times U(1)_{\text{inst}}$.

Our first goal is to ensure that we have a single field $X$ with charges $(+,-,0)$ localized on $\Sigma$.  The number of fields of each type that we obtain on $\Sigma$ is given by
\begin{equation}\label{matter}
\begin{split}
n_{+-0}&= h^0\left(\Sigma,K_{\Sigma}^{1/2}\otimes V_1|_{\Sigma}\otimes V_2^{-1}|_{\Sigma}\right)\\
n_{-+0}&= h^0\left(\Sigma,K_{\Sigma}^{1/2}\otimes V_1^{-1}|_{\Sigma}\otimes V_2|_{\Sigma}\right)\,,
\end{split}
\end{equation}
where $V_i|_{\Sigma}$ is the restriction of line bundle $V_i$ to $\Sigma.$

Next, we turn to fermi zero modes arising from \textquotedblleft 3-7\textquotedblright strings connecting the D3-instanton to the $D7_1$ on $S_1$.  These are counted by
\begin{equation}\begin{split}
n_{+0-}&= h^1\left(S_1,V_1\otimes V_{\text{inst}}^{-1}\right)=-\chi\left(S_1,V_1\otimes V_{\text{inst}}^{-1}\right)\\
n_{-0+}&= h^1\left(S_1,V_1^{-1}\otimes V_{\text{inst}}\right)=-\chi\left(S_1,V_1^{-1}\otimes V_{\text{inst}}\right)\,.
\end{split}\end{equation}

Finally, we consider fermi zero modes arising from \textquotedblleft 3-7\textquotedblright strings connecting the D3-instanton to the $D7_2$ on $S_2$ which intersects $S_1$ along $\Sigma$.  These are counted by
\begin{equation}\begin{split}
n_{0+-}&= h^0\left(\Sigma,K_{\Sigma}^{1/2}\otimes V_2|_{\Sigma}\otimes V_{\text{inst}}^{-1}|_{\Sigma}\right)\\
n_{0-+}&= h^0\left(\Sigma,K_{\Sigma}^{1/2}\otimes V_2|_{\Sigma}^{-1}\otimes V_{\text{inst}}|_{\Sigma}\right)\,.
\end{split}\end{equation}
If we can choose bundles so that
\begin{equation}
n_{+-0}=n_{-0+}=n_{0+-}=1\qquad\text{and}\qquad n_{-+0}=n_{+0-}=n_{0-+}=0
\label{Polconsts}\end{equation}
then we will obtain a single chiral multiplet $X$ and two fermi zero modes $\alpha$ and $\beta$ connecting the D3-instanton to matter branes with charges
\begin{center}
\begin{tabular}{c|c|c|c}
Field & $U(1)_1$ & $U(1)_2$ & $U(1)_{\text{inst}}$ \\ \hline
$X$ & $+$ & $-$ & $0$ \\
$\alpha$ & $0$ & $+$ & $-$ \\
$\beta$ & $-$ & $0$ & $+$
\end{tabular}
\end{center}
This will allow a coupling of the form
\begin{equation}\alpha\beta X\label{PabXcoupling}\end{equation}
to appear in the instanton action which is needed in order to generate the superpotential \eqref{WPinst}.

We must also consider bosonic $D3$-$D7_1$ and $D7_1$-$D3$ zero modes, though.  These are counted as
\begin{equation}
S_-'\otimes \Bigl(H^0(S_1,V_1\otimes V_\text{inst}^{-1})\oplus H^0(S_1,V_1^{-1}\otimes V_{\text{inst}})\Bigr) \,,
\end{equation}
where $S_{-}'$ is anti-chiral spinor in $\mathbb{R}^4$.
We further require that $V_1\otimes V_{\text{inst}}^{-1}$ be a non-trivial supersymmetric
bundle on $S_1$ so that there are no bosonic zero modes in this sector.
Finally, there are no bosonic $D3$-$D7_2$ and $D7_2$-$D3$ zero modes since
there is a non-vanishing zero point energy in the NS sector for intersections with two ND directions.

\subsubsection{Instanton Action and Generating the Polonyi Term}

With the above conditions and including both \eqref{P33couplings} and \eqref{PabXcoupling}, we obtain an instanton action of the form
\be\label{instaction}
S_{\text{inst}}=t_{S_1}+ \psi^{\alpha}\theta_{\alpha} + \alpha \beta X +
\mu_{\dot \alpha}\Bigl({\bar b}_{KK}^{\dot \alpha}f_{KK}+b_{KK}^{\dot \alpha}{\bar f}_{KK}\Bigr)+
M_{b,KK} {\bar b}_{KK}b_{KK}+M_{f,KK} {\bar f}_{KK}f_{KK}\,,
\ee
where $\psi^{\alpha}$ is fermion super-partner of
\be \label{chirtPol}
t_{S_1}=\int_{S_1} {(J+4\pi^2\alpha' f_{\text{inst}})^2\over 4\pi^2 g_s}+i S_{WZW}^{inst} \,,
\ee
where $f_{\text{inst}}=c_1(V_{\text{inst}})$. Here $J$ is the K\"ahler form on $S_1,$ while $S_{WZW}^{inst}$
is the WZW coupling to RR fields which we will discuss in detail in Section \ref{sec:Axions}.
The superpotential generated by this instanton is obtained from the
two-point correlation function in the instanton background
\be
\langle\psi_1^{\dg}\psi_2^{\dg} \rangle \sim \partial^2_t W_{\text{inst}} \,,
\ee
which results in
\be \label{polsup}
\ba
W_{\text{inst}}\sim &e^{-t_{S_1}} \int d\alpha\, d\beta\, d\mu\,  df_{KK}\, db_{KK}\, d{\bar f}_{KK}\, d{\bar b}_{KK}\cr
& \qquad \qquad \qquad
e^{-\alpha \beta X-\mu_{\dot \alpha}\Bigl({\bar b}_{KK}^{\dot \alpha}f_{KK}+b_{KK}^{\dot \alpha}{\bar f}_{KK}\Bigr)
-M_{b,KK} {\bar b}_{KK}b_{KK}-M_{f,KK} {\bar f}_{KK}f_{KK}} \,.
\ea
\ee
Now we note that the integral over the anti-chiral $3$-$3$ fermi zero mode $\mu_{\dot{\alpha}}$ can be done and only affects the measure of integration for
the KK modes\footnote{In the models where there are 3-7 zero bosonic and 7-3 zero  fermionic modes
$b_0,{\bar b}_0$ and $f_0,{\bar f}_0$ extra to the modes we have,
the integral over $\mu$ gives measure for zero modes $f_0,b_0,{\bar f}_0,{\bar b}_0.$}
 $b_{KK},{\bar b}_{KK}$ and $f_{KK},{\bar f}_{KK}$.  As such, the superpotential is indeed non-zero and is given by
\be
W_{\text{inst}}=F_X X \,,
\ee
where the coefficient $F_X$ is given by
\begin{equation} \label{FX}
F_X = M^{2}_{S_{1}} \cdot q \cdot w^{\nu(f_{inst})} \,.
\end{equation}
In the above, $M_{S_{1}}$ denotes the characteristic mass scale for D3-instantons wrapping $S_{1}$,
$q = \text{exp}(2\pi i \tau_{S_{1}})$ is the universal contribution from the 1-instanton sector, and $w = \text{exp}(2 \pi i \tau_{IIB})$ is the contribution from the non-trivial $U(1)$ flux with instanton number $\nu(f_{inst})$ on the worldvolume of the Euclidean 3-brane. An appropriately complexified $\tau_{S_{1}}={it_{S_1}\over 2\pi}\vert_{f_{inst}=0}$ is the holomorphic gauge coupling of the 4d gauge theory, and $\tau_{IIB} = c_{0} + i/g_{s}$ denotes the axio-dilaton.  Here, we are crudely treating $\tau_{IIB}$ as an overall constant, although in more general configurations $\tau_{IIB}$ can be position dependent, and the contribution from $w$ should then be viewed as a rough averaging over the instanton density with the axio-dilaton. Finally, as expected, the energy scale $\sqrt{F_{X}}$ is exponentially suppressed in comparison to $M_{S_{1}}$.

\subsubsection{Leading Order Behavior of the Instanton Expansion}

It is important to note that this model will only correspond to a Polonyi-like model if no higher powers $X^{m}$ for $m>1$
appear in the 1-instanton contribution to the superpotential.
In subsection \ref{sec:OneInst} we show all such higher order terms indeed vanish.
Further, we demonstrate in subsection \ref{sec:kInst} that higher powers are generated in the $k$-instanton
sector with $k>1,$ so that the total D3-instanton generated superpotential has the form
\be
W_{\text{inst}}^{tot}= \sum_{m,k,f_{inst,k,m}} (c_{m,k,f_{inst}} q^{k} \cdot w^{\nu(f_{inst,k,m})}) X^{m} = C + F_XX + O(e^{-2t_{S_1}})  \,.
\ee
In the above, the sum on $k$ denotes the contribution from the $k$-instanton sector, and the sum on $f_{inst,k,m}$ denotes the contribution from internal fluxes on the D3-instanton with instanton number $\nu(f_{inst,k,m})$.  The contribution given by $C$ is independent of $X$ and can be treated as a relatively unimportant constant shift when the volume modulus of the del Pezzo is stabilized due to high scale dynamics. In addition, the coefficient of the leading order contribution to the Polonyi-like term is given by equation \eqref{FX}.   We find that all non-zero internal fluxes which contribute in the $k = 1$, $m = 1$ sector have the same instanton number. Moreover, as indicated by the second equality, we find that $c_{m,1,f_{inst}}=0$ for $m>1$ and all possible internal fluxes. Note, however, that due to the large suppression factors, the  extra terms will have a negligible effect on physics near the SUSY-breaking minimum.


\subsection{
The 1-Instanton Sector}
\label{sec:OneInst}

In this subsection we determine the net contribution to the effective superpotential from the 1-instanton sector.  This involves summing over all ``trivializable'' internal fluxes on the worldvolume theory defined by the D3-instanton in the presence of a fixed background flux on the 7-branes.

The background fluxes on the 7-branes are partially fixed by the requirement that a single 4d chiral superfield $X$
localizes on the matter curve $\Sigma$.  For simplicity, we shall assume that $\Sigma$ is a genus zero curve.  In
this case, a single zero mode localizes on $\Sigma$ when the supersymmetric line bundles $V_{i}$ satisfy the condition:
\be \label{convv}
V_2|_{\Sigma}=V_1|_{\Sigma}\otimes \mathcal{O}(-1) \,.
\ee
If we impose this condition then we are still free to adjust $V_1$.  With this assumption, we now determine all combinations of line bundles $V_1$ and $V_{\text{inst}}$ which can contribute to the 4d superpotential.
In particular, we will show that a linear term in $X$ is generated, and that all higher powers of the form
$X^m$ for $m>1$ are indeed absent.

As a first step, we introduce the line bundle associated with open strings connecting the D3-instanton to the D7$_{1}$-brane
\be
\mathcal{L} \equiv V_{\text{inst}}\otimes V_1^{-1} \,.
\ee
The conditions \eqref{Polconsts} to have the right
number of fermi zero modes  to produce a term $W_{\text{inst}}\sim X^m$
amount to the following three constraints on $\mathcal{L}$:
\be \label{condl}
\chi(S_1,\mathcal{L})=-m,\qquad
\chi(S_1,\mathcal{L}^{-1})=0,\qquad
\mathcal{L}\vert_{\Sigma}=\mathcal{O}(-m-1) \,.
\ee
These conditions can be recast as
\be \label{tocompare}
c_1^2(\mathcal{L})=-m-2,\qquad
K_{S_1}\cdot c_1(\mathcal{L})=m,\qquad
[\Sigma] \cdot c_1(\mathcal{L})=-m-1 \,.
\ee
To proceed further, we now require $S_1=dP_M, \, S_2=dP_N$ with $M,N\ge 2$ and
fix a class for the genus zero curve $\Sigma$.

For concreteness, we specify $\Sigma$ as a rigid divisor in the homology class
\be
[ \Sigma ]=H-E_1-E_2  \in H_2(S_1,\mathbb{Z}) \,.
\ee
Most generally, the line bundle $\mathcal{L}$ can be written as
\be
\mathcal{L}=\mathcal{O}\Bigl(b_0H+\sum_{i=1}^M b_iE_i\Bigr),\quad b_0,b_i \in \mathbb{Z} \,,
\ee
so that the conditions (\ref{tocompare}) become explicitly
\be \label{expl}
b_0^2-\sum_{j=1}^M b_j^2=-m-2\,,\qquad
 3b_0+\sum_{j=1}^Mb_j=-m\,,\qquad
 b_0+b_1+b_2=-m-1 \,.
\ee
When $M=2$, we note that the above conditions imply $b_0 = 1/2$, so that there are no integral solutions in this case.
We therefore restrict our choice of del Pezzos to
\be
S_1=dP_M\,, \qquad
S_2=dP_N \qquad N\ge 2,\quad M\ge 3 \,.
\ee
To set notation, we denote:
\be\label{xydef}
x =\sum_{j=4}^Mb_j\,,\qquad
y=\sum_{j=4}^Mb_j^2 \qquad M\ge 4 \,,
\ee
while $x=y=0$ for $M=3$. Solving (\ref{expl}) for $b_1$, $b_2$ and $b_3$ yields

\be
b_1=-\half \Bigl((b_0+m+1)\mp \sqrt{Q}\Bigr)\,,\qquad
b_2=-\half \Bigl((b_0+m+1)\pm \sqrt{Q}\Bigr)\,,\qquad b_3=1-2b_0-x \,,
\ee
where in the above, the $\mp$ and $\pm$ are correlated, and
\be
Q=3-m^2-2(1-x-2b_0)^2-2(m+1)b_0+b_0^2-2y \,.
\ee
Note that $\mathcal{L}$ corresponds to a solution only if $b_2$ is an integer,
which in particular implies $Q\ge 0$.

In Appendix C we demonstrate that even without specifying a specific polarization for the K\"ahler form,
for $M=3,4$ there are no solutions of (\ref{expl}) with $m\ge 2$.  We also show that
for $M=3,4$ all solutions with $m=1$ have the form{\footnote{Note that in general such a flux cannot be ``trivializable'' so the simplest way to engineer a Polonyi model is to simply set $V_1^{-1}={\cal{O}}(E_p-E_1-E_2)$.  In that case, summing over ``trivializable'' fluxes $V_{\text{inst}}$ we will find that only the case of trivial flux contributes to the superpotential.}}
\be \label{result}
\mathcal{L}=\mathcal{O}(E_p-E_1-E_2)\,,\qquad p\ne 1\,,  2 \,.
\ee
Determining all solutions to (\ref{expl}) is
somewhat more delicate for $M \geq 5$ because
the variables $x$ and $y$ now depend on more than one parameter.  Nevertheless, the analysis remains tractable for an
appropriate polarization of the K\"ahler form.  Specifying the K\"ahler class $J$ as in (\ref{kahler}), we find that for $m=1$ all supersymmetric solutions have the form (\ref{result}), and more crucially, there
are no supersymmetric solutions of (\ref{expl}) for $m\ge 2.$  In particular, this establishes that the
only $X$ dependent contribution from the 1-instanton sector to the superpotential is given by the Polonyi term.


\subsection{Constraints From Axion Shifts}
\label{sec:Axions}

Though this successfully demonstrates that the sum over the internal flux sectors of the D3-instanton wrapping $S_1$ generates a linear superpotential \eqref{WPinst}, we now perform a consistency check on this result.  As in subsection \ref{6dINTERP}, the point is that regardless of the background fluxes on the D7's, the 6d gauge theory on $\mathbb{R}^{4}\times \Sigma$ is anomalous.  The Green-Schwarz mechanism cancels the corresponding anomaly through an appropriate axion-like coupling.  In particular, this implies that we should always expect an appropriate shift in the RR potentials of the theory.  To this end, we now study the shifts of various RR axions under the anomalous $U(1)$'s in the model.  This will also make it clear that a single D3-instanton can never simultaneously generate both a linear term and higher degree terms in the superpotential.

To see how this comes about in the explicit Polonyi-like model we have found, recall that the theory has two $U(1)$ gauge group factors
associated with the two D7's wrapped on the del Pezzo surfaces $S_1$ and $S_2$.  Under $U(1)_1\times U(1)_2$ gauge transformations, the gauge
field transforms as:
\be
A_{i} \mapsto A_{i}+d\lambda_{i}\quad i=1,2\,,
\ee
while the bifundamental field $X$ living on $\Sigma$ transforms as{\footnote{We normalize the gauge fields so that
\be
D_{\mu}X=\partial_{\mu} X -2\pi i (A_1-A_2)X \,,
\ee
in order to simplify our discussion of anomalies below, i.e. to be able
to write the Chern character of a line bundle $V_i$ as $ch(V_i)=e^{F_i}.$
This, in particular, implies that the kinetic term in the Lagrangian for gauge fields
is $-\half F\wedge *F$.}}
\be
X\mapsto e^{2\pi i(\lambda_1-\lambda_2)}X\,.
\ee
As described in detail in \cite{Minasian:1997mm,Green:1996dd,Cheung:1997az}, $U(1)_1\times U(1)_2$ transformations also cause the RR potentials $C_{RR}$ to undergo a shift.  This is important because these fields appear in the WZW coupling $S^{\text{inst}}_{WZW}$ of the instanton suppression factor $e^{-t}$
\be
t \equiv \int_{S_1}{\Bigl(J+4\pi^2\alpha' c_1(V_{\text{inst}})\Bigr)^2\over 4\pi^2 g_s}+iS_{WZW}^{inst} \,,
\ee
so that their shift effectively causes the quantity $e^{-t}$ to carry its own $U(1)_1\times U(1)_2$ charge.  If we properly account for this charge, any superpotential coupling must be fully invariant with respect to the corresponding gauge symmetry.  This means in particular that a necessary condition for the generation of a linear superpotential of the form $e^{-t}X$ is that the WZW coupling $S_{WZW}^{inst}$ shifts as
\be \label{odin}
S_{WZW}^{inst}\mapsto S_{WZW}^{inst} + 2\pi(\lambda_1-\lambda_2) \,.
\ee
In what follows, we set $4\pi^2 \alpha'=1$ below to simplify various formulae.

To find the shifts of various axions in the background of
intersecting D7-branes we first recall the anomaly on their intersection, which is usually referred to as the $I_{(1,2)}$-brane in the literature.  The
worldvolume ${\bf I}_{(1,2)}$ of the $I_{(1,2)}$-brane is defined as the
intersection of the worldvolumes of the two D7-branes, i.e.
${\bf I}_{(1,2)}=\mathbb{R}^4\times \Sigma$.
For $i=1,2$ let $g_i$ be an embedding of the i-th D-brane worldvolume $W_i$ into ten
dimensional spacetime.
Let $NW_i$ be the normal bundle to the $i$-th D-brane, which is isomorphic to $K_{S_i}$.
The anomalous variation on ${\bf I}_{(1,2)}$ is
given by\footnote{We use the normalization of RR fields as in \cite{Polchinski:1998rq} with $\alpha=1.$} \cite{Minasian:1997mm, Green:1996dd, Cheung:1997az}
\be \label{anomvar}
\delta_{gauge} S_{I-brane}^{(1,2)}=2\pi \int_{\mathbb{R}^4\times \Sigma} X^{(1)}\,,\qquad
X=2e^{(F_1-F_2)}{\hat A}(T{\bf I}) \,,
\ee
where the cohomology class of the gauge field-strength is half-integral
$[F_i]=c_1(V_i)+\half K_{S_i}$ as derived in \cite{Minasian:1997mm} and
extended to backgrounds with non-trivial NSNS flux in \cite{Freed:1999vc}.
This is commonly referred to as the Freed-Witten anomaly.

In \eqref{anomvar}, $X^{(1)}$ is computed by the descent procedure
\be
X=1+dX^{(0)},\quad \delta_{gauge} X^{(0)}=dX^{(1)} \,.
\ee
and ${\hat A}$ is the A-roof genus
\be
{\hat A}=1-{p_1\over 24}+ \cdots \,.
\ee
The ambiguity in the choice of $X^{(0)}$ is fixed by requiring that it be symmetric between the branes
\be
X^{(0)}=\Bigl(Y_1{\tilde Y}_2+{\tilde Y}_1Y_2\Bigr)^{(0)} \,,
\ee
where
\be
\label{defy}
Y_i=e^{F_i}{\sqrt{\hat A(TW_i)}\over\sqrt{\hat A(NW_i)}}\,,\qquad
{\tilde Y}_i=e^{-F_i}{\sqrt{\hat A(TW_i)}\over\sqrt{\hat A(NW_i)}} \,.
\ee
This leads to
\be \label{anomvarii}
\delta_{gauge} S_{I-brane}^{(1,2)}=2\pi \int_{\mathbb{R}^4\times \Sigma} \Bigl(Y_1{\tilde Y}_2+{\tilde Y}_1Y_2\Bigr)^{(1)} \,.
\ee

The variation (\ref{anomvarii}) is canceled by
anomaly inflow due to anomalous WZW
couplings to RR fields,  in the presence of D7-branes.
More concretely, the WZW coupling to the $k$-th D7-brane  is given by
\be \label{axdef}
S_{WZW}^{(k)}=\mu \int_{\mathbb{R}^4\times
S_k}\Bigl(C_{RR}+Y^{(0)}_kF_{RR}\Bigr)\,,\qquad k=1,2 \,.
\ee
Here $C_{RR}$ ($F_{RR}$) is a formal sum of all RR gauge fields (field strengths). In the presence
of D7-branes, $F_{RR}$ satisfies the modified Bianchi identity \cite{Minasian:1997mm, Green:1996dd, Cheung:1997az}
\be \label{bimod}
dF_{RR}=\mu \sum_{j=1}^2 {\tilde Y}_j\delta(W_j)\,.
\ee
Meanwhile, $C_{RR}$ transforms
non-trivially under the $U(1)_1 \times U(1)_2$ gauge transformations as:
\be
\label{crrmod} \delta C_{RR}=-\mu  \sum_{j=1}^2 {\tilde
Y}^{(1)}_j\delta(W_j).
\ee
As a result of (\ref{bimod}) and (\ref{crrmod}), the WZW couplings $S_{WZW}^{(k)}$ transform as
\be
\sum_{k=1}^2 \delta_{gauge} S_{WZW}^{(k)}=-\mu^2\int_{\mathbb{R}^4\times \Sigma}
\Bigl(Y_1{\tilde Y}_2+{\tilde Y}_1Y_2\Bigr)^{(1)} \,,
\ee
and cancel the anomalous variation on the $I_{(1,2)}$-brane
(\ref{anomvar}) for $\mu^2=2\pi$.

Next suppose that a D3-instanton wraps $S_1$. There are now three I-branes
with worldvolumes
\be{\bf I}_{(1,2)}=\mathbb{R}^4\times \Sigma,\quad
{\bf I}_{(1,inst)}=S_1,\quad {\bf I}_{(2,inst)}=\Sigma.
\ee
Thus, the total anomaly under $U(1)_1\times U(1)_2\times U(1)_{\text{inst}}$ is
\be
\delta_{gauge} S_{I-brane}^{(1,2)}+\delta_{gauge} S_{I-brane}^{(1,inst)}+
\delta_{gauge} S_{I-brane}^{(2,inst)} \,,
\ee
where $\delta_{gauge} S_{I-brane}^{(1,2)}$ is given by (\ref{anomvarii})
and the other two summands are similar
\be
\ba \label{summands}
\delta_{gauge}S_{I-brane}^{(1,inst)}&=2\pi\int_{S_1}K_{S_1}\wedge \Bigl(Y_1{\tilde Y}_{\text{inst}}+{\tilde Y}_1Y_{\text{inst}}\Bigr)^{(1)} \cr
\delta_{gauge}S_{I-brane}^{(2,inst)}&=2\pi\int_{\Sigma}\Bigl(Y_2{\tilde Y}_{\text{inst}}+{\tilde Y}_2Y_{\text{inst}}\Bigr)^{(1)} \,,
\ea
\ee
where we have used the property \cite{Cheung:1997az}
\be
\delta(S_1)\delta(S_1)=K_{S_1}\delta(S_1) \,.
\ee

As we discussed above, $\delta_{gauge} S_{I-brane}^{(1,2)}$ is canceled by an anomalous transformation
of the WZW couplings on the D7-branes. Meanwhile, the anomalous transformation
of the WZW couplings on the D3-brane cancels the remaining two anomalies
\be \label{axionshift}
\delta_{gauge}S_{WZW}^{inst}=-\Bigl(\delta_{gauge}S_{I-brane}^{(1,inst)}+
\delta_{gauge}S_{I-brane}^{(2,inst)}\Bigr)\,.
\ee

Because of the extra factor of $K_{S_1}$ in \eqref{summands}, the evaluation of \eqref{axionshift} requires us to expand various forms only up to degree two:
\be
\ba
Y_i{\tilde Y}_{\text{inst}} &=1+(F_i-F_{\text{inst}})
\cr
{\tilde Y}_iY_{\text{inst}}&=1+(F_{\text{inst}}-F_i)
\,,
\ea
\ee
where
\be
F_i=c_1(V_i)+\half K_{S_i}\,,\qquad F_{\text{inst}}=c_1(V_{\text{inst}})+\half K_{S_i} \,,
\ee
so that $F_i-F_{\text{inst}}=c_1\left(V_i\otimes V_{\text{inst}}^{-1}\right).$
Substituting these expansions into (\ref{axionshift})
yields \be
\label{fininstii}\delta_{gauge}
S_{WZW}^{inst}=
-2\pi (\lambda_{\text{inst}}-\lambda_2)c_1\left(V_{\text{inst}}\otimes V_2^{-1}\right)\cdot \Sigma -2\pi  (\lambda_{\text{inst}}-\lambda_1)c_1
\left(V_{\text{inst}}\otimes V_1^{-1}\right)\cdot K_{S_1}.
\ee
Requiring now that $S_{WZW}^{inst}$ transform as in (\ref{odin}) leads to the following necessary conditions on the bundles
\be
c_1\left(V_{\text{inst}}\otimes V_2^{-1}\right)\cdot \Sigma=-1,\quad
c_1 \left(V_{\text{inst}}\otimes V_1^{-1}\right)\cdot K_{S_1}=1 \,.
\ee
To compare this with the previously obtained zero mode conditions of line (\ref{tocompare}),
we rewrite the necessary conditions  in terms of $\mathcal{L}=V_{\text{inst}}\otimes V_1^{-1}$:
\be
K_{S_1}\cdot \mathcal{L}=1\,,\qquad
\Sigma \cdot c_1(\mathcal{L})+ \Sigma \cdot
\Bigl(c_1(V_1)-c_1(V_2)\Bigr)=-1 \,.
\ee
Now recall that we choose D7
bundles to satisfy (\ref{convv}) so that the constraints obtained from the
anomalous transformation of the WZW coupling reproduces the two linear
equations for $m=1$ in  (\ref{tocompare}), but misses the quadratic
one. This is what we expected since the zero mode counting should give
stronger constraints.


\subsection{$k$-Instanton Sector}
\label{sec:kInst}
\subsubsection{Zero Mode Content}

We finally turn our attention to multi-instanton configurations.  While we shall find
that these multi-instantons can generate superpotential couplings of the form $X^m$
with $m>1$ the corresponding suppression factors will always be quite small.

To study multi-instantons, we consider $k$ D3-branes wrapped on $S_1$  and switch on a supersymmetric
vector bundle $V_{\text{inst}}$ with structure group $U(1)\times H$ with $H
\subset SU(k)$. Such a bundle can be defined in terms of a stable vector
bundle $V_{{inst},H}$ with structure group $H$ and supersymmetric line bundle $\mathcal{V}.$
We further assume, as in Section 2.2, that $\mathcal{V}$ is trivial in the three-fold $B.$
Let $G$ be the commutant of $H$ in $SU(k)$ and decompose the adjoint $U(k)$
bundle under the $G\times U(1)\times  H$ subgroup as
\be
Adj_{U(k)}=\mathcal{O}
\oplus \left(Adj_G,\mathcal{O} \right)
\oplus \left(1_G,V_{inst,H}\otimes V_{inst,H}^{\ast}\right)
\oplus \bigoplus_s \left( \left(r_s,V_{{\cal R}_s}\right) \oplus\left({\bar r}_s,V_{\bar {\cal R}_s}\right)\right) \,,
\ee
where $1_G$ is a singlet under $G$ and $r_s$ is a non-trivial irreducible representation of $G$.
Meanwhile, if $V_{inst,H}$ is non-trivial, then
$V_{{\cal R}_s}$ is a supersymmetric vector bundle on $S_1$ in an irreducible representation ${\cal R}_s$
of $H$.

The various fermion zero modes are counted in Appendix \ref{sec:appendix33}.
The D3-D3 fermion zero modes are sections of
\be
\ba
&\quad S_+' \otimes \Biggl(H^{0}\left(S_1, K_{S_1}\otimes Adj_{U(k)}\right) \oplus H^{2}\left(S_1, K_{S_1}\otimes Adj_{U(k)}\right)
\oplus H^{1}\left(S_1, Adj_{U(k)}\right)\Biggr)\quad  \cr
\oplus
&\quad S_-' \otimes \Biggl(H^{0}\left(S_1, Adj_{U(k)}\right)\oplus H^{2}\left(S_1, Adj_{U(k)}\right)\oplus
H^{1}\left(S_1, K_{S_1}\otimes Adj_{U(k)}\right)\Biggr) \,,
\ea
\ee
where $S_+'(S_-')$ denotes the (anti-)chiral spin bundle in $\mathbb{R}^4$.

We assume that $V_{inst,H}$ is a non-trivial supersymmetric bundle. Utilizing the
vanishing theorem for supersymmetric bundles on del Pezzo surfaces (see for example \cite{Beasley:2008dc}) it follows that the D3-D3 fermion zero modes are given by:
\be
\ba
\theta &\in S_+'\otimes H^2(S_1,K_{S_1}) \cr
\mu &\in S_-'\otimes H^0(S_1,\mathcal{O}) \cr
\zeta &\in Adj_G \otimes S_+'\otimes H^2(S_1,K_{S_1})\cr
 \xi &\in Adj_G \otimes S_-'\otimes H^0(S_1,\mathcal{O})\cr
\lambda_s &\in r_s \otimes S_+' \otimes H^1(S_1,V_{{\cal R}_s}) \cr
 {\bar \lambda}_s & \in {\bar r}_s \otimes S_+' \otimes H^1(S_1,V_{\bar {\cal R}_s})\cr
\eta_s & \in r_s \otimes S_-' \otimes H^1(S_1,K_{S_1}\otimes V_{{\cal R}_s})\cr
{\bar \eta}_s &\in {\bar r}_s \otimes S_-' \otimes H^1(S_1,K_{S_1}\otimes V_{\bar {\cal R}_s}) \cr
\lambda_{1_G} &\in 1_G \otimes S_+' \otimes H^1(S_1,V_{inst,H}\otimes V_{inst,H}^{\ast})\cr
\eta_{1_G} &\in 1_G \otimes S_-' \otimes H^1(S_1,K_{S_1}\otimes V_{inst,H}\otimes V_{inst,H}^{\ast}) \,.
\ea
\ee

Since $h^0(S_1,\mathcal{O})=h^2(S_1,K_{S_1})=1$ the D3-D3 fermion zero mode content is
\begin{itemize}
\item 1-copy of $\theta, \, \mu, \,\zeta,\, \xi$
\item $n_{\lambda_s}$ copies of
$\lambda_s$ and ${\bar \eta}_s$
\item  $n_{\eta_s}$ copies of $\eta_{s}$ and ${\bar \lambda}_s$
\item $n_{1_G}$ copies of $\lambda_{1_G}$ and $\eta_{1,G}$
\end{itemize}
Here
\be
\ba
n_{\lambda_s} &=h^1(S_1,V_{{\cal R}_s})=-\chi(S_1,V_{{\cal R}_s}) \cr
n_{\eta_s}&=h^1(S_1,V_{{\cal R}_s}^{\ast})=-\chi(S_1,V_{{\cal R}_s}^{\ast}) \cr
n_{1_G}&=h^1(S_1,V_{inst,H}\otimes V_{inst,H}^{\ast}) \,.
\ea
\ee
Note that when $H$ is a trivial subgroup of $SU(k)$ so that $G=SU(k),$ there are no $\lambda$ and
$\eta$ modes.

In addition to the fermion modes, we also find D3-D3 bosons which are sections of
\be
T\mathbb{R}^4 \otimes \Omega^{0,0}\otimes Adj_{U(k)} \,,
\ee
where $T\mathbb{R}^4$ is tangent bundle to $\mathbb{R}^4$. So that the D3-D3 bosonic zero
modes are:
\be
x_0^{\mu},\quad Y^{\mu}_a,\quad a=1,\ldots,dim(G) \,,
\ee
where we used that  $H^0(S_1,V_{{\cal R}_a})=0$ for a non-trivial
supersymmetric bundle $V_{{\cal R}_a}$.
We will also need the KK modes $Z_{KK}^{\mu}$ which arise from non-zero modes of
the Laplacian acting on $\Omega^{0,0}\otimes 1_G\otimes
V_{inst,H}\otimes V_{inst,H}^{\ast}$. The coupling of these modes
with zero mode fermions $\lambda_{1_G}$ and $\eta_{1_G}$ is important
to obtain a non-vanishing path-integral in our computation of the superpotential below.

To count D3-D7 (D7-D3) modes we first decompose the vector bundle in fundamental
representation of $U(k)$ under $G\times U(1)\times H$ as
\be
V_{fund}=(r_{fund,G},\mathcal{V}) \oplus(1_G,V_{inst,H}\otimes \mathcal{V} ) \,.
\ee

Let us denote ${\cal E}=V_{inst,H}\otimes \mathcal{V}\otimes V_1^{-1}$ and
$\mathcal{L}=\mathcal{V}\otimes V_1^{-1}$. Recall that we require
$\Sigma=\mathbb{P}^1$ and $V_2|_{\Sigma}\otimes
V_1^{-1}|_{\Sigma}=\mathcal{O}(-1)$  in order to have a single
chiral field $X$ on $\Sigma.$

Then, the D3-D7 (D7-D3) fermion zero modes $\alpha, \, \delta  \in
r_{\overline{fund},G}$ and $\beta,\, \gamma \in r_{fund,G}$ are counted
as
\be
\ba
\#_{\alpha} &=h^0\Bigl(\mathbb{P}^1,\mathcal{O}(-2)\otimes \mathcal{L}^{-1}|_{\Sigma}\Bigr) \cr
\#_{\gamma}&=h^0\Bigl(\mathbb{P}^1, \mathcal{L}|_{\Sigma} \Bigr) \cr
\#_{\beta}&=-\chi\Bigl(S_1,\mathcal{L}\Bigr) \cr
\#_{\delta} &=-\chi\Bigl(S_1,\mathcal{L}^{-1}\Bigr)\,,
\ea
\ee
where in counting the $\beta$ and $\delta$ zero modes we have again used the vanishing
theorem for supersymmetric vector bundles on del Pezzo surfaces \cite{Beasley:2008dc}.
Note that when $H=SU(k)$ i.e. $G$ is trivial there are no $\alpha,\, \beta, \, \gamma,\, \delta$
modes.

Meanwhile, the D3-D7 (D7-D3) fermion zero modes $\talpha, \, \tdelta, \,
\tbeta,\, \tgamma \in 1_G$ are counted as
\be
\ba
\#_{\talpha} &=h^0\Bigl(\mathbb{P}^1,\mathcal{O}(-2)\otimes {\E^{\ast}}|_{\Sigma}\Bigr) \cr
\#_{\tgamma}&=h^0\Bigl(\mathbb{P}^1, \E|_{\Sigma} \Bigr) \cr
\#_{\tbeta}&=-\chi\Bigl(S_1,\E\Bigr) \cr
\#_{\tdelta} &=-\chi\Bigl(S_1,{\E^{\ast}}\Bigr) \,.
\ea
\ee
Note that when $H$ is a trivial subgroup of $SU(k)$ i.e. $G=SU(k),$ there are no $\talpha,\, \tbeta, \, \tgamma,\, \tdelta$
modes.

The D3-D7 (D7-D3) bosonic zero modes are counted as
\be
\ba
&S_-' \otimes \Bigl((1_G,\ H^0(S_1, {\E^{\ast}})) \oplus (r_{\overline{fund},G} ,\ H^0(S_1, \mathcal{L}^{-1}))\Bigr)\quad  \cr
\oplus
&S_-' \otimes \Bigl((1_G,\ H^0(S_1,\E)) \oplus (r_{fund,G},\ H^0(S_1, \mathcal{L}))\Bigr) \,,
\ea
\ee
i.e. they are absent for non-trivial supersymmetric bundles $\E$ and $\mathcal{L}.$

As before, in order to compute the net contribution to the superpotential, we must also include the coupling of the zero modes found above
to D3-D7 and D7-D3 KK modes.
Let us call $b_{KK}^{\dot \alpha}$ and ${\bar b}_{KK}^{\dot \alpha}$ the KK bosonic modes which arise from non-zero modes of the Laplacian acting on
$S_-'\otimes \Omega^{0,0} \otimes \mathcal{L}$ and
$S_-'\otimes \Omega^{0,0} \otimes \mathcal{L}^{-1}$, respectively.
Also, let $f_{KK}$ and ${\bar f}_{KK}$ be the KK fermi modes which correspondingly arise from $\Omega^{0,2} \otimes K_{S} \otimes \mathcal{L}$
 and $\Omega^{0,2} \otimes K_{S} \otimes \mathcal{L}^{-1}$.
There are other KK modes but we do not specify them since all fermi zero modes in our
computation of the correlator $\langle\psi_1^{\dg}\psi_2^{\dg} \rangle$ below
are saturated without considering them.

\subsubsection{Generating $X^m$, $ m\ge 1$}

Let us require that we have a single copy of $\beta \in r_{fund,G}$ and
$\alpha \in r_{\overline{fund},G},$ as well as $Q$-copies of
$\talpha \in 1_G$ and $\tbeta \in 1_G$ and no other 3-7 or 7-3 fermion zero modes.
We show below that integrating over these zero modes yields the term $W_{\text{inst}}\sim X^{g+Q}$ with $g=dim\left(r_{fund,G}\right).$

In order to generate this interaction term, $\mathcal{L}$ must satisfy the constraints of equation (\ref{condl}) for $m=1$
previously found for the 1-instanton sector.  In addition, the bundle $\E$ must satisfy the constraints:
\be \label{constronE} \chi(S_1,\E)=-Q,\quad
\chi(S_1,{\E^{\ast}})=0,\quad \E|_{\Sigma}=\sum_{p > 0}
\mathcal{O}(-p-1)^{\oplus l_p} \oplus \mathcal{O}(-1)^{\oplus (rk-d)} \,,
\ee
where we have introduced the partitions $Q=\sum_p p\, l_p$ and $d=\sum_p l_p$, and $rk$ is the dimension of the fundamental representation of $H$. We discuss the solutions of these constraints in the next subsection.

Now we proceed to generate $W_{\text{inst}}.$
The action for $k$ D3-branes  has the form (contraction of indices in the unbroken gauge group $G$ is implicit)
\be
\ba
S_{\text{inst}}
&=k(t_{S_1}+ \psi^{\alpha}\theta_{\alpha}) +
 \alpha \cdot \beta X + \sum_{u=1}^Q \talpha^{(u)} \tbeta^{(u)} X+
\mu^{\dot \alpha}\Bigl(b_{KK\, \dot \alpha}{\bar f}_{KK} +{\bar b}_{KK\, \dot \alpha}f_{KK} \Bigr)+
M_{b,KK} {\bar b}_{KK}b_{KK}\cr
&+M_{f,KK} {\bar f}_{KK}f_{KK}+Tr \xi^{\dot \beta}[{\hat Y}_{{\dot \beta} \alpha},\zeta^{\alpha}]
+ \sum_s \sum_{v=1}^{n_{\lambda_s}}
{\bar \eta}_s^{(v)\, \dot \beta} {\hat Y}^{a}_{{\dot \beta} \alpha} T_a^{(r_s)}\lambda_s^{(v)\, \alpha}+
\sum_s \sum_{w=1}^{n_{\eta_s}}\eta_s^{(w)\, \dot \beta} {\hat Y}^{a}_{{\dot \beta} \alpha} T_a^{({\bar r}_s)}{\bar \lambda}_s^{(w)\, \alpha}\cr
&+
\sum_{l=1}^{n_{1_G}}\eta_{1_G}^{(l)\, \dot \beta} {\hat Z}^{KK}_{{\dot \beta} \alpha} \lambda_{1_G}^{(l)\, \alpha}
+M_{Z,KK} {Z}_{\mu \, KK}\, Z^{\mu}_{KK}+\sum_{c=1}^{dim(G)}\Biggl( -\half D_c^2+ D_c {\bar \eta}^c_{\mu \nu} [Y^{\mu},Y^{\nu}]\Biggr) \,,
\ea
\ee
where
\be
\ba
{\hat Y}_{{\dot \beta} \alpha} &={\hat Y}^a_{{\dot \beta} \alpha }T_a^{(Adj_G)}\cr
{\hat Y}^a_{{\dot \beta} \alpha }&=Y^a_{\mu}\sigma^{\mu}_{{\dot \beta} \alpha} \cr
{\hat Z}^{KK}_{{\dot \beta} \alpha }&=Z^{KK}_{\mu}\sigma^{\mu}_{{\dot \beta} \alpha} \,,
\ea
\ee
 and
$\psi^{\alpha}$ is the fermion super-partner of $t_{S_1}$ defined in
(\ref{chirtPol}). Finally, the $T_a^{(r_s)}$ denote generators of the group $G$ in the representation $r_s$
and ${\bar \eta}^c_{\mu \nu}$ denotes the t'Hooft symbol.

The integral over $D$ and $\xi,$ as well as $\eta_s,{\bar \eta}_s$ defines the integration measure on the
instanton moduli space
\be
\ba
{\cal M}(Y,\zeta,\lambda_s,{\bar \lambda}_s)
&=\epsilon_{\dot  \alpha \dot \beta} \prod_{a=1}^{dim(G)} f_{abc} f_{ade}\Bigl({\hat Y}_b \zeta_c\Bigr)_{\dot \alpha}\Bigl({\hat Y}_d \zeta_e\Bigr)_{\dot \beta} \cr
&\times
\prod_s \epsilon_{\dot  \gamma \dot \delta} \prod_{I_s=1}^{dim(r_s)}
\Bigl({\hat Y}^{a_1} T_{a_1}^{(r_s)}\lambda_s\Bigr)_{I_s \, \dot \gamma}
\Bigl({\hat Y}^{b_1} T_{b_1}^{(r_s)}\lambda_s\Bigr)_{I_s \, \dot \delta}\cr
&\times \prod_s \epsilon_{\dot  \rho \dot \epsilon} \prod_{{\bar I}_s=1}^{dim({\bar r}_s)}
\Bigl({\hat Y}^{a_2} T_{a_2}^{({\bar r}_s)}{\bar \lambda}_s\Bigr)_{{\bar I}_s \, \dot \rho}
\Bigl({\hat Y}^{b_2} T_{b_2}^{({\bar r}_s)}{\bar \lambda}_s\Bigr)_{{\bar I}_s \, \dot \epsilon}\times
\int [dD] e^{-S(D,Y)}
\ea
\ee
with
\be
S(D,Y)=\sum_{c=1}^{dim(G)}\Bigl( -\half D_c^2+ D_c {\bar \eta}^c_{\mu \nu} [Y^{\mu},Y^{\nu}]\Bigr)\,.
\ee
Note that, similar to the 1-instanton sector, the
couplings
$$\mu^{\dot \beta}\Bigl( {\bar b}_{KK \, \dot \beta} f_{KK}+ b_{KK \, \dot \beta} {\bar f}_{KK}\Bigr)+
\sum_{l=1}^{n_{1_G}}\eta_{1_G}^{(l)\, \dot \beta} {\hat Z}^{KK}_{{\dot \beta} \alpha} \lambda_{1_G}^{(l)\, \alpha}$$
are crucial for obtaining a non-vanishing contribution from the path integral. Integrating over $\mu$ gives a modified measure factor
for the KK modes $b_{KK},{\bar b}_{KK}$ and $f_{KK},{\bar f}_{KK}$, while integrating over $\eta_{1_G},\, \lambda_{1_G}$ gives a modified measure factor
for $Z_{KK}.$

The superpotential generated by instantons is obtained from the
two-point correlation function in the instanton background
\be
\langle\psi_1^{\dg}\psi_2^{\dg} \rangle \sim \partial^2_{t_{S_1}} W_{\text{inst}}
\,,
\ee
which results in
\be \label{polsupmulti} W_{\text{inst}}=\Upsilon_{KK}
e^{-kt_{S_1}} \int d\alpha\, d\beta \, d\talpha\, d\tbeta \,  dY \, d\zeta \, d\lambda \,
{\cal M}(Y,\zeta,\lambda,{\bar \lambda})\,
 e^{-\alpha \cdot \beta X-\sum_{u=1}^Q \talpha^{(u)} \tbeta^{(u)} X}\,
\sim e^{-kt_{S_1}}\,X^{m}
\ee
where
\be
m=g+Q,\quad g=dim(r_{fund,G})\le k,\qquad Q=\sum_{p > 0} pl_p\,, \qquad \sum_{p > 0} l_p \le dim({\cal R}_{fund,H}) \,.
\ee
The integral over KK modes with the measure factor discussed above is included in the prefactor $\Upsilon_{KK}$.
Note that for $k\ge 2$ both cases $m>k$ and $1\le m \le k$ are a priori possible.
For example, $g=0$ for $H=SU(k)$  but $Q$ could be larger than k. It would be interesting to determine whether
there exist stable vector bundles $\E$ solving (\ref{constronE}) for $Q>k$.


\subsubsection{Sample Solutions of the Constraints}

The simplest example when $k$ D3-instantons generate $W\sim X^m$ is to
take $H$ to be a trivial subgroup of $SU(k)$ so that $G=SU(k).$  In this case,
there are no modes of the type $\talpha,\tbeta.$
Further, specifying the line bundle $\mathcal{L}$ as in subsection \ref{sec:OneInst}, we find a single copy of
$\alpha$ and $\beta$ which respectively transform in the anti-fundamental and fundamental representation of
$U(k)$. Therefore, $g=k$ in this case and $Q=0$, which implies $m=k.$  Technically speaking,
however, this contribution is already accounted for in the exponentiation of the 1-instanton sector.

We now present a contribution to the 2-instanton sector which cannot be obtained by exponentiating the
contributions from the 1-instanton sector.  To this end, consider a configuration with $k=2$ and $H=SU(2),$ i.e.  $g=0$.
Our strategy will be to construct a rank 2 bundle obtained from an extension of two supersymmetric line bundles. As we now show,
the multi-instanton sector also contributes to the linear superpotential term $W\sim X$. In this case, $Q=1$ which corresponds to the following conditions
on a supersymmetric $U(2)$ vector bundle $\E$
\be \label{constronEiii}
\chi(S_1,\E)=-1\,,\qquad
\chi(S_1,{\E^{\ast}})=0\,,\qquad
\E|_{\Sigma}=\mathcal{O}(-2)\oplus \mathcal{O}(-1) \,.
\ee
A stable rank two irreducible vector bundle $\E$ may be constructed as an extension of
a supersymmetric line bundle $T=\mathcal{O}(E_3-E_1)$ by a supersymmetric line bundle
$\mathcal{L}_p=\mathcal{O}(E_p-E_1-E_2)$ for $p\ne 3.$ The restriction on $p$
comes from the requirement that\footnote{To compute this cohomology group, we invoke the vanishing
theorem for del Pezzo surfaces, and $\chi(S_1,T^*\otimes \mathcal{L}_p)=-1$, $ p\ne 3.$ }
$h^{1}(S_1, T^*\otimes \mathcal{L}_p)=1,$ in order to obtain a non-trivial extension.
Also note that we assume that line bundle $T$ is trivializable in the three-fold $B.$
This is required since, as we discussed in section 2, only instantons with $c_1(V_{inst})$
trivial in $B$  can contribute to superpotential.

Assuming that as in subsection \ref{sec:OneInst} the class of the matter curve $[\Sigma]=H-E_1-E_2$, we have
$T^*\otimes\mathcal{L}_p\vert_\Sigma=\mathcal{O}(-1)$ so that $h^1(\mathbb{P}^1,\mathcal{O}(-1))=0$.  In particular,
this implies that over $\Sigma$ the extension trivializes\footnote{Although any $U(2)$ vector bundle will restrict on $\mathbb{P}^1$ to
a sum of two line bundles, to establish that this decomposition is the same as that given by the sum of the restrictions $\mathcal{L}_p|_{\Sigma}$
and $T|_{\Sigma}$ requires showing that the extension trivializes over $\Sigma$.}
$$\E\vert_{\Sigma}=\mathcal{L}_p|_{\Sigma}+T|_{\Sigma}=\mathcal{O}(-2)+\mathcal{O}(-1),$$
so that the third condition in (\ref{constronEiii}) is satisfied.

Further, since $\E$ is an extension, we have
\be
ch(\E)=ch(\mathcal{L}_{p})+ch(T)\,.
\ee
Using also
 $$ \chi(S_1,\mathcal{L}_p)=-1,\quad \chi(S_1,\mathcal{L}_p^{-1})=0$$
and
$$\chi(S_1,T)=0$$ we compute
\be
\chi(S_1,\E)=-1,\quad  \chi(S_1,{\E^{\ast}})=0 \,.
\ee
Thus, the first two conditions on $\E$ in (\ref{constronEiii}) are also satisfied.

Let us comment on the stability of the vector bundle $\E$.
Since $\E$ is obtained as a non-trivial extension
of the supersymmetric line bundle $T$ by the supersymmetric line bundle $\mathcal{L}_p,$
the stability of $\E$ follows if we restrict the choice of the K\"ahler form $J$
so that
\be
c_1(T)\cdot J < c_1(\mathcal{L}_p)\cdot J \,.
\ee
Using our ansatz for $J$ as in (\ref{kahler}) we find that the stability of
$\E$ requires
\be
B_2-B_p+B_3 >0 \,.
\ee



\section{D3-Instantons and Pure $SU(N)$ Gauge Theory}

\label{sec:puresun} In this Section we study the low energy limit of $SU(N)$
gauge theory defined by a stack of 7-branes wrapping a rigid divisor $S$ in
the 3-fold base of an F-theory compactification. As in the rest of this
paper, for simplicity we assume $S$ is a del Pezzo surface. Although the
zero mode content of the corresponding 4d effective theory is the same as 4d
pure $\mathcal{N} = 1$ $SU(N)$ gauge theory, the low energy dynamics of the
two systems are not identical.

The first distinction between the two theories comes from the fact that in
the 7-brane theory, the holomorphic gauge coupling constant:
\begin{equation}
\tau_{YM}=\frac{4\pi i}{g_{YM}^{2}}+\frac{\theta_{YM}}{2\pi}
\end{equation}
actually originates from a normalizable mode. Now, in a full string
compactification, all coupling constants of the 4d effective theory may be
viewed as the vevs of fields which are typically non-dynamical after taking
an appropriate decoupling limit. In the present case, however, the del Pezzo
surface has dimension greater than the middle dimension of the 3-fold base $%
B $ of the F-theory compactification. This implies that even in a local
model, the volume modulus corresponds to a normalizable mode. For most
purposes, it is sufficient to assume that some other dynamics at energy
scales far above that set by the gauge theory stabilizes this modulus.
Nevertheless, in this Section we shall not assume that this modulus is
frozen to a background value.

The second distinction between the 8d gauge theory and the 4d effective
theory is that the $\tau_{YM}$ dependent terms of the superpotential will in
general receive additional contributions beyond those expected from purely
4d considerations. Indeed, we have already seen one example of this kind in
Section \ref{sec:polonyi} where D3-instantons with non-trivial internal
worldvolume flux played a crucial role in the analysis. In particular, the
superpotential of the Polonyi-like model contains a contribution independent
of the field $X$. While in that context we assumed that this term could be
treated as a constant, here we shall study a similar contribution to the
pure $SU(N)$ gauge theory in its own right.

The rest of this Section is organized as follows. \ By way of comparison, we
first review gaugino condensation in 4d $SU(N)$ gauge theory, and its
proposed role in moduli stabilization. \ Next, we show that the 1-instanton
sector of the 7-brane theory generates an important correction to the usual
4d glueball superpotential.

\subsection{Review of 4d Gaugino Condensation and Moduli Stabilization}

\label{review}

To frame the discussion to follow, first recall that at low energies, 4d $%
\mathcal{N}=1$ $SU(N)$ gauge theory confines, undergoing gaugino
condensation. \ The effective superpotential for the glueball fields is \cite%
{Veneziano:1982ah}:%
\begin{equation}
W_{VY}(S_{g},\tau _{YM})=2\pi i\tau _{YM}S_{g}-N\left( S_{g}\log \frac{S_{g}%
}{\Lambda _{0}^{3}}-S_{g}\right)   \label{WVY}
\end{equation}%
where $\Lambda _{0}$ denotes the UV cutoff of the gauge theory, and
\begin{equation}
S_{g}=-\frac{1}{32\pi ^{2}}Tr_{SU(N)}W^{\alpha }W_{\alpha }
\end{equation}%
denotes the glueball field of the confining gauge theory. \ The critical
points of $W_{VY}$ with respect to $S_{g}$ are then given by:%
\begin{equation}
\frac{S_{g}}{\Lambda _{0}^{3}}=\zeta _{N}\exp \left( 2\pi i\tau
_{YM}/N\right) \equiv \zeta _{N}q^{1/N}\,,  \label{SCRIT}
\end{equation}%
where $\zeta _{N}$ is an $N^{th}$ root of unity and $q=\text{exp}(2\pi i\tau
_{YM})$. Fractional instanton effects break the $U(1)$ R-symmetry down to $%
\mathbb{Z}_{2N}$, which is further broken to to $\mathbb{Z}_{2}$ once $S_{g}$
develops a non-zero vev. Indeed, shifting $\theta _{YM}$ by $2\pi $
cyclically permutes the confining vacua of the theory.

As explained near the beginning of this Section, even in a limit where
gravity has been decoupled, the volume modulus of the del Pezzo surface
remains dynamical. When the effective superpotential is given by equation (%
\ref{WVY}), the F-term equations of motion for $S_{g}$ and $\tau_{YM}$ yield:%
\begin{align}
\frac{\partial W_{VY}}{\partial S_{g}} & =2\pi i\tau_{YM}-N\log\frac{S_{g}}{%
\Lambda_{0}^{3}}=0 \\
\frac{\partial W_{VY}}{\partial\tau_{YM}} & =2\pi iS_{g}=0
\end{align}
so that the vacuum exhibits runaway behavior to $\text{Im}%
(\tau_{YM})\rightarrow \infty$.

This runaway behavior can be avoided when this same modulus behaves as the
gauge coupling constant in another theory, or through some other moduli
stabilization mechanism. In the literature, gaugino condensation from
7-branes has been employed for precisely this purpose. See \cite%
{Denef:2005mm} for an explicit application of this type, as well as \cite%
{Denef:2008wq} for a recent review on this and related matters. From the
perspective of the 4d theory, these effects can typically be encapsulated in
terms of an additional contribution to the effective superpotential:
\begin{equation}  \label{WSTAB}
W_{eff}(S_{g},\tau_{YM})=W_{VY}(S_{g},\tau_{YM}) + h\left(\tau_{YM}\right)
\text{.}
\end{equation}
where $h$ serves to stabilize the modulus $\tau_{YM}$. In the next
subsection, we show that even in the context of a local model, $h$ receives
a contribution linear in $q$.

\subsection{Isolated 7-Branes and Instantons}

In this subsection we show that D3-instantons contribute a $\tau_{YM}$
dependent term to the 4d superpotential which in the purely 4d theory would
be regarded as a constant. After showing that in an appropriate regime of
parameters the F-term equations of motion can stabilize the volume modulus
of the divisor wrapped by the 7-brane, we briefly comment on potential
applications for constructing models of supersymmetry breaking.

Although our ultimate interest is in the low energy dynamics of the 7-brane
theory, we can avoid all subtleties pertaining to strong coupling effects
associated with glueball fields by computing possible \textquotedblleft
constant\textquotedblright \ term contributions to the 4d superpotential
above the scale of confinement. Our first aim is to show that D3-instantons
can contribute terms which are independent of the glueball fields of the 4d
effective theory. In fact, as we now argue, such terms are absent in the
purely 4d theory. To see how this comes about, first consider D3-instanton
contributions which admit an interpretation as instantons in the purely 4d
theory. In contrast to 4d $\mathcal{N} = 1$ $U(1)$ gauge theory, 4d pure $%
\mathcal{N} = 1$ $SU(N)$ gauge theory certainly admits non-trivial instanton
configurations. To see what sort of F-term contributions are in principle
possible, recall that a 4d gauge theory instanton corresponds to the special
case where the internal fluxes of the D3-instanton and 7-brane exactly
align. In all cases where the D3-instanton forms a bound state with the
7-brane, the $\mu_{\dot{\alpha}}$ universal zero mode integral is typically
saturated via interactions with the 3-7 and 7-3 strings. When a non-zero
contribution to the superpotential is possible, these additional factors
simply contribute a modified measure factor. In the present case, there is a
single bosonic and fermionic zero mode in both the 3-7 and 7-3 sector
because $h^{0}(S,\mathcal{O}) = h^{2}(S,K_S) = 1$. Now, this zero mode
corresponds to a bifundamental under the gauge group $U(1)_{\text{inst}}
\times SU(N)$. With notation as before, we therefore let $b_{a}^{\dot{\alpha}%
}$ and $\overline{b}_{a}^{\dot{\alpha}}$ denote these bosonic modes and $%
f^{a}$ and $\overline{f}^{a}$ denote these fermionic modes. Here, the index $%
a = 1,...,N$. As usual, these modes interact with $\mu_{\dot{\alpha}}$
through the couplings:
\begin{equation}
\mu_{\dot \alpha}\Bigl(b_{a}^{\dot \alpha}{\bar f}^{a}+{\bar b}_{a}^{\dot
\alpha}f^{a}\Bigr) \,.
\end{equation}
Integrating over the $\mu_{\dot \alpha}$ zero modes, the resulting measure
factor contains terms quadratic in the fermionic zero modes $f^{a}$ and $%
\overline{f}^{a}$. While it is therefore possible to saturate one pair of
fermionic zero mode integrals, this will still leave $N-1$ unsaturated terms
so that when $N > 1$, this contribution will vanish. Note that a priori this
contribution will not vanish when an explicit mass term is included in the
instanton action, as would happen if these were instead KK modes rather than
zero modes. Typically, these additional fermionic integrations can be
saturated through interactions with the 7-7 modes of the 7-brane theory.
These could in principle correspond either to terms involving $%
Tr_{SU(N)}W^{\alpha}W_{\alpha}$, or KK modes of the 7-brane theory. It would
be interesting to characterize the form of such contributions which are
specific to 7-branes.

It follows from the discussion above that a \textquotedblleft
constant\textquotedblright\ contribution from a D3-instanton is possible
only when neither 3-7 nor 7-3 zero modes are present. We count these modes
in Appendix B.2.2, and find that they are absent if
\begin{equation}
H^{i}(S,V_{\text{inst}})=H^{i}(S,V_{\text{inst}}^{-1})=0  \label{vanish}
\end{equation}%
for all $i$. In the above, $V_{\text{inst}}$ denotes the line bundle which
has support on the D3-instanton worldvolume. In \cite{Beasley:2008kw}, line
bundles satisfying equation \eqref{vanish} were classified for del Pezzo $k$
surfaces. It follows from this result that $V_{\text{inst}}=\mathcal{O}%
_{S}(\alpha )$, where $\alpha $ denotes a root of the Lie algebra $E_{k}$
viewed as an element of $H_{2}(S,\mathbb{Z})$.

In order for the Euclidean D3-brane configuration to contribute to the
superpotential, the corresponding flux threading the D3-brane must be
trivializable in the threefold base $B$. Note that this is consistent with
the fact that $c_{1}(V_{\text{inst}})\cdot K_{S}=0$.

We now explain why this D3-instanton configuration contributes to the 4d
effective superpotential. As throughout this paper, the KK modes of the 3-7
and 7-3 strings play a crucial role in the analysis. Let $b_{KK}^{\dot
\alpha}$ and ${\bar b}_{KK}^{\dot \alpha}$ denote bosonic KK modes in $%
S_-^{\prime 0,0} \otimes V_{\text{inst}}$ and $S_-^{\prime 0,0} \otimes V_{%
\text{inst}}^{-1}$ respectively. Similarly, let $f_{KK}$ and ${\bar f}_{KK}$
denote fermionic KK modes from $\Omega^{0,2} \otimes K_{S} \otimes V_{\text{%
inst}}$ and $\Omega^{0,2} \otimes K_{S} \otimes V_{\text{inst}}^{-1}$
correspondingly. There are other KK modes which we can neglect for the
purposes of this discussion because all fermi zero modes in our computation
of the superpotential will already be saturated without considering any
couplings to these unspecified KK modes.

Next consider the $3$-$3$ zero mode content. Because the worldvolume theory
of the D3-instanton is a $U(1)$ gauge theory, the adjoint bundle on the
D3-instanton worldvolume is trivial. Hence the only $3$-$3$ fermion zero
modes (see Appendix B.2.1 for counting) comprise a chiral spinor $%
\theta_{\alpha}$ and anti-chiral spinor $\mu_{\dot \alpha}$. As explained
previously in Section \ref{sec:generalities}, there is a coupling in the
instanton action
\begin{equation}
\mu_{\dot \alpha}\Bigl(b_{KK}^{\dot \alpha}{\bar f}_{KK}+{\bar b}_{KK}^{\dot
\alpha}f_{KK}\Bigr) \,,
\end{equation}
so that the integral over $\mu_{\dot \alpha}$ yields a modified measure
factor for the KK modes.

The bosonic $3$-$3$ collective coordinates yield the measure factor $d^{4}x$
over $\mathbb{R}^4$, while the fermionic $3$-$3$ zero modes $\theta_{\alpha}$
yield the $d^{2}\theta$ measure factor over the fermionic coordinates of
superspace. \ The instanton contribution to the superpotential is then:%
\begin{equation}
W_{inst}=-\Lambda_{0}^{3}\cdot \lambda q\text{,}
\end{equation}
where $\lambda$ is determined by a moduli dependent worldvolume determinant
factor, as well as an axio-dilaton dependent factor which depends on 
the internal flux on the D3-instanton.

The 4d effective superpotential for the isolated 7-brane theory with gauge
group $SU(N)$ is therefore:%
\begin{equation}
W_{eff}(S_{g},\tau _{YM})=2\pi i\tau _{YM}S_{g}-N\left( S_{g}\log \frac{S_{g}%
}{\Lambda _{0}^{3}}-S_{g}\right) -\Lambda _{0}^{3}\cdot \lambda q\text{.}
\label{WEFFINST}
\end{equation}%
An important consequence of the above is that the F-term equations of motion
with respect to $\tau _{YM}$ now yield the relation:
\begin{equation}
\frac{S_{g}}{\Lambda _{0}^{3}}=\lambda q\text{.}
\end{equation}%
Note that in contrast to the $S_{g}$ equation of motion expected based on
the 4d glueball superpotential, the $\tau _{YM}$ equation of motion would
suggest that the $U(1)$ R-symmetry of the $SU(N)$ gauge theory is broken
completely.

In appropriate circumstances, the D3-instanton contribution may in fact
stabilize the volume of the del Pezzo surface. Indeed, the critical points
satisfy the relations:
\begin{align}
\frac{S_{g}}{\Lambda_{0}^{3}} & = \zeta_{N} q^{1/N} \\
\frac{S_{g}}{\Lambda_{0}^{3}} & =\lambda q\text{.}
\end{align}
One class of solutions corresponds to the runaway solution defined by $q
\rightarrow 0$ and $S_g \rightarrow 0$. On the other hand, there is another
branch of supersymmetric vacua given by:
\begin{align}
\tau_{YM} & =\frac{i}{2\pi}\frac{N}{N-1}\log\lambda \\
\frac{S_g}{\Lambda_{0}^{3}} & =\frac{1}{\lambda^{1/(N-1)}}\text{,}
\end{align}
where to reduce notational clutter we have suppressed the explicit
dependence on various roots of unity. In particular, we see that for
sufficiently large values of $\lambda$, this vacuum solution is well-defined.

Note, however, that at smaller values of $\lambda$, this branch of solutions
becomes ill-defined. Indeed, when $|\lambda| < 1$, $g_{YM}^{2}<0$ and the
effective field theory description appears to break down. For example, the
glueball field $S_g$ formally becomes of order $\Lambda_{0}^{3}$. It is in
principle possible that strong coupling effects could trigger the analogue
of a geometric transition, although there are well-known obstructions to
this for K\"{a}hler surfaces in Calabi-Yau 4-folds. A perhaps simpler
possibility is that supersymmetry is broken in the regime of small $\lambda$.%
\footnote{%
We note that this is not in conflict with a count of the number of vacua
obtained from the Witten index. Indeed, it is likely that whatever the
dynamics may be, additional massless modes will enter the spectrum near this
region of parameter space, so that the Witten index will likely jump.} It
would clearly be of interest to engineer explicit models of supersymmetry
breaking which exploit the presence of this linear term in $q$.



\section*{Acknowledgements}

We thank C. Beasley, M. Buican, S. Franco, D. Jafferis, S. Kachru, D.R. Morrison, H. Ooguri, Y.
Ookouchi, D. Robbins, K. Saraikin,  S. Sethi and T. Weigand for valuable discussions. The
work of JJH and CV is supported in part by NSF grants PHY-0244821
and DMS-0244464. The research of JJH is also supported by an NSF
Graduate Fellowship. The work of JM and SSN was supported by John A.
McCone Postdoctoral Fellowships. The work of NS was supported in
part by the DOE-grant DE-FG03-92-ER40701. JJH, JM, and CV thank the
Simons Center for Geometry and Physics at Stony Brook and the 2008
Simons Workshop in Mathematics and Physics, which directly led to
this collaboration, for kind hospitality.  JJH also thanks the 2008
Amsterdam Summer Workshop on String Theory for hospitality while
some of this work was performed. JM also thanks the SITP at Stanford
University and the Aspen Center for Physics for their hospitality
during the course of this work. SSN thanks the IPMU and the
University of Tokyo for kind hospitality.

\newpage

\newpage

\setcounter{section}{0}
\setcounter{subsection}{0}


\appendix

\section{Useful Formulae}
\label{app:Useful}

In this Appendix we list a few useful formulae.  As explained  earlier, the complex surfaces under consideration are always chosen to be del Pezzo $dP_n$. Here, $dP_n$ denotes the surface $\mathbb{P}^{2}$ blown up at $n = 0,...,8$ points in general position.  A basis of divisors is given by
$H$ and $E_i$, $i=1,\cdots,n$, with the intersections
\be\label{dPBasis}
H \cdot H =1 \,,\qquad
E_i\cdot E_j = -\delta_{ij} \,,\qquad
H\cdot E_i =0 \,.
\ee
The canonical class is
\be\label{dPKlass}
K_{dP_n} = - 3 H + \sum_{i=1}^n E_i \,.
\ee
Given a curve in a class $[\Sigma]$, the genus $g$ of the curve is given by the relation
\be
[\Sigma] \cdot ([\Sigma] + \mathcal{K}_{dP_n}) = 2 g -2 \,.
\ee

The bundles we consider are required to be supersymmetric
\be
\label{SusyBundle}
\int_{dP_n} c_1(\mathcal{L})\wedge J=0 \,,
\ee
where $J$ is a chosen polarization, i.e. a family of K\"ahler-forms which includes the large volume limit.
By abuse of notation, we shall typically refer to the K\"ahler class by the same name.  One choice
of K\"ahler class was given in \cite{Beasley:2008dc}
\be \label{choicei}
J=AH+\sum_{i=1}^n B_iE_i\ee
with $A\gg 1$, $B_i < 0$ and $|B_i| \sim O(1)$ for $i = 1,...,M$


When computing bundle cohomologies, the following relations are useful
\be\label{PoneCoho}
h^0\left(\mathbb{P}^1, \mathcal{O}(k) \right) =
    \left\{
        \begin{array}{lcc}
        k>0:& & k+1 \cr
        k=0:& & 1 \cr
        k<0:& & 0
        \end{array}
    \right.
\ee
\begin{equation}\chi(dP_{n},{\cal{V}})=1-\frac{1}{2}K_{dP_{n}}\cdot c_1({\cal{V}})+\frac{1}{2}\left[c_1({\cal{V}})\right]^2 \text{.}\end{equation}


\section{Zero-Modes in the Presence of D-Brane Instantons}\label{app:ZeroModes}

In this Appendix, we describe the counting of zero modes that is needed to determine the possible contributions of D-brane instantons.  We begin by reviewing the zero mode counting of \cite{Beasley:2008dc} which determines the spectrum of modes that arise in intersecting 7-brane configurations.  This is done in two ways.  First, we implement the twisting of \cite{Beasley:2008dc}.  Second, we adopt a more naive counting based on open strings which is valid only when we have a perturbative description.  We will find that, when both are applicable, the two methods yield identical results.

We will then turn our attention to counting instanton fermion zero modes, again adopting both procedures when applicable.
\subsection{Zero Mode Counting in the Twisted SYM Theory}

\subsubsection{7-Brane Zero Modes}\label{app:77}

Let us first review the counting of zero modes on a 7-brane wrapping the surface $S$ obtained in \cite{Beasley:2008dc}.  The spectrum of the 7-brane worldvolume theory is obtained by dimensionally reducing the 10-dimensional maximally supersymmetric Yang-Mills theory down to 8-dimensions.  The fermion consists of a single chiral spinor in the $\mathbf{16}_+$ which in turn decomposes into an $\mathbf{8}_+\oplus \mathbf{8}_-$.  Because the $\mathbf{16}_+$ is real, the $\mathbf{8}_-$ should be viewed as the conjugate of the $\mathbf{8}_+$.

We will twist by a $U(1)_J$ which lies inside $U(2)\subset SO(4)$ so let us decompose these spinors under $SO(7,1)\times U(1)_R\rightarrow SO(3,1)\times U(2)\times U(1)_R$
\begin{equation}\begin{split}\mathbf{8}_+&\rightarrow \left[(\mathbf{2},\mathbf{1}),\mathbf{2}_0,+\frac{1}{2}\right]\oplus\left[(\mathbf{1},\mathbf{2}),(\mathbf{1}_{+1}\oplus\mathbf{1}_{-1}),+\frac{1}{2}\right]\\
\mathbf{8}_-&\rightarrow\left[(\mathbf{2},\mathbf{1}),(\mathbf{1}_{+1}\oplus\mathbf{1}_{-1}),-\frac{1}{2}\right]\oplus\left[(\mathbf{1},\mathbf{2}),\mathbf{2}_0,-\frac{1}{2}\right] \,.
\end{split}
\end{equation}
In this expression the $SO(3,1)$ representations are specified by giving the corresponding $SU(2)\times SU(2)$ content while for $U(2)$ representations carry a subscript indicating their charge with respect to the diagonal $U(1)_J$ subgroup.  Twisting with $J_{top}=J\pm 2R$ leads to fermions which transform as
\begin{equation}\begin{split}
\mathbf{8}_+&\rightarrow \left[(\mathbf{2},\mathbf{1})\otimes \mathbf{2}_{+1}\right]\oplus\left[(\mathbf{1},\mathbf{2})\otimes \left(\mathbf{1}_{+2}\oplus\mathbf{1}_0\right)\right]\\
\mathbf{8}_- &\rightarrow \left[(\mathbf{1},\mathbf{2})\otimes\mathbf{2}_{-1}\right]\oplus\left[(\mathbf{2},\mathbf{1})\otimes(\mathbf{1}_0\oplus\mathbf{1}_{-2})\right] \,.
\end{split}\end{equation}
Letting $\Omega^{1}$ denote the holomorphic cotangent bundle on $S$, this leads to twisted fermions
\begin{equation}\begin{split}\bar{\eta}_{\dot{\alpha}}&\in\Gamma(\text{ad}(P))\\
\psi_{\alpha}&\in\Gamma(\overline{\Omega}^{1}\otimes\text{ad}(P))\\
\bar{\chi}_{\dot{\alpha}}&\in\Gamma(\overline{\Omega}^{2}\otimes\text{ad}(P))
\end{split}\end{equation}
from the $\mathbf{8}_+$ where we have allowed for the possibility of a nontrivial gauge bundle $\text{ad}(P)$.  The fields coming from the $\mathbf{8}_-$ are simply conjugates of these and give nothing new.  Zero modes are now counted by the corresponding bundle cohomologies
\begin{equation}\label{TestEq}
\bar{\eta}_{\dot{\alpha}}\in H^0(S,\text{ad}(P))\qquad\psi_{\alpha}\in H^1(S,\text{ad}(P))\qquad \bar{\chi}_{\dot{\alpha}}\in H^2(S,\text{ad}(P)) \,.
\end{equation}
In the absence of a bundle which breaks the bulk gauge group, we can display the full zero mode content on the 7-brane as
\begin{equation}
\left[S'_-\otimes H^0(S,\text{ad}(P))\right]\oplus\left[S_+'\otimes H^1(S,\text{ad}(P))\right]\oplus\left[S_-'\otimes H^2(S,\text{ad}(P))\right]\,,
\end{equation}
where $S'_{\pm}$ are chiral/anti-chiral spin bundles in $\mathbb{R}^{3,1}$.  Recalling that CPT conjugation simply dualizes the cohomology group, we can also write this as
\begin{equation}S_+'\otimes\left[H^0(S,K_{S}\otimes\text{ad}(P))\oplus H^1(S,\text{ad}(P))\oplus H^2(S,K_{S}\otimes\text{ad}(P))\right]\,,
\label{fermadj}\end{equation}
where we have applied Serre duality
\begin{equation}
H^p(S,{\cal{V}})=H^{2-p}(S,K_{S}\otimes {\cal{V}}^{\ast})^{\ast}\,.
\end{equation}
Of course, in addition to \eqref{fermadj} we get the conjugate fermions as well.

Now, let us suppose that the bulk gauge group is broken by fluxes.  In addition to fields in the adjoint of the unbroken gauge group, decomposing the adjoint representation of the bulk gauge into irreducible representations of the unbroken gauge group and the group in which the instanton takes values, there will also be fields in the bifundamental representation{\footnote{By this we mean bifundamental in the sense described in \cite{Beasley:2008dc}.} $R$ and, when it is a complex representation, its conjugate $\bar{R}$.  Denoting the bundle associated to $R$ as $V_R$, we obtain chiral fermions charged under $R$ from
\begin{equation}
S_+'\otimes\left[H^0(S,K_{S}\otimes V_R)\oplus H^1(S,V_R)\oplus H^2(S,K_{S}\otimes V_R)\right]\,.
\end{equation}
In the case of D7-branes, we can think of these as coming from open strings of a specified orientation (derived from the Chan-Paton indices) from the 7-brane to itself.  When $R$ is complex, we also get chiral fermions charged under $\bar{R}$ from
\begin{equation}
S_+'\otimes\left[H^0(S,K_{S}\otimes V_R^{\ast})\oplus H^1(S,V_R^{\ast})\oplus H^2(S,K_{S}\otimes V_R^{\ast})\right]\,,
\end{equation}
which in the D7-brane case corresponds to open strings of the opposite orientation.

\subsubsection{Zero Modes from Matter Curves}\label{app:77p}

We now recall the counting of $7-7'$ modes localized at the intersection of a 7-brane and a second $7'$-brane along a matter curve $\Sigma$.  These modes live on a 6-dimensional defect.  The fermion arises from a  hypermultiplet transforming in the bifundamental representation $R$ and hence corresponds to 6-dimensional spinor in the $\mathbf{4}'$ of $SO(5,1)$.  In particular, it is uncharged under the $SU(2)_R$ symmetry and so is unaffected by the twisting, which uses a $U(1)_R\subset SU(2)_R$.  This means that to count the modes we simply need to decompose the $\mathbf{4}'$ under $SO(5,1)\rightarrow SO(3,1)\times U(1)$
\begin{equation}
\mathbf{4'}\rightarrow \left((\mathbf{2},\mathbf{1}),-\frac{1}{2}\right)\oplus\left((\mathbf{1},\mathbf{2}),+\frac{1}{2}\right) \,.
\end{equation}
These correspond in turn to a 4d chiral and a 4d anti-chiral spinor
\begin{equation}
\lambda_{\alpha}\in S_+'\otimes H^0(\Sigma,K_{\Sigma}^{1/2}\otimes V_R)\qquad \bar{\lambda}^c_{\dot{\alpha}}\in S'_-\otimes H^1(\Sigma,K_{\Sigma}^{1/2}\otimes V_R) \,,
\end{equation}
where $V_R$ is the bundle corresponding to the associated bifundamental representation.  We can again express the matter content purely in terms of $SO(3,1)$ chiral spinors by conjugating $\bar{\lambda}^c$ to obtain a chiral spinor in the representation $\bar{R}$.  This means that we in fact have 4d chiral fermions in the representation $R$ from
\begin{equation}
S_+'\otimes H^0(\Sigma,K_{\Sigma}^{1/2}\otimes V_R)
\end{equation}
and 4d chiral fermions in the representation $\bar{R}$ from
\begin{equation}
S_+'\otimes H^0(\Sigma,K_{\Sigma}^{1/2}\otimes V_R^{\ast})^{\ast}\,.
\end{equation}
In the case of D7-branes, these can be thought of as arising from open strings connecting the D7- and D7'-branes with opposite orientations.

\subsubsection{D3-Instanton: 3-3 Zero Modes}\label{app:33}
\label{sec:appendix33}

Let us now turn to bulk zero modes of a D3-instanton, which we refer to as $3$-$3$ modes.  Again we start with the $\mathbf{16}_+$ from maximally supersymmetric Yang-Mills in 10-dimensions and decompose it under $SO(10)\rightarrow SO(4)\times SO(4)\times U(1)_R$.  The only difference now is that the interpretation of the spacetime $SO(4)$ corresponds to a global symmetry of the D3-instanton theory.  This has no effect on the twisting, which only involves the second $SO(4)$ factor and the $U(1)_R$ factor of the symmetry algebra.  As such, the twisting proceeds exactly as in subsection \ref{app:77}, meaning that we have $3$-$3$ zero modes coming from
\begin{equation}
S_+'\otimes\left[H^0(S,K_{S}\otimes\text{ad}(P))\oplus H^1(S,\text{ad}(P))\oplus H^2(S,K_{S}\otimes\text{ad}(P))\right] \,.
\end{equation}
and their conjugates.  Note that for a single D3-instanton with gauge group $U(1)$,
turning on a nontrivial gauge bundle does not alter the $3$-$3$ zero mode content.
Finally, we note that when counting the number of complex fermi zero modes we should
include an extra factor of two due to the fact that the bundle $S_+'$ is complex
and has rank two.

\subsubsection{D3-Instanton: 3-7 and 3-7' Zero Modes}\label{app:37}

We now study zero modes which arise from strings connecting the D3-instanton to various 7-branes.  In particular, we consider 3-7 modes, which connect the instanton wrapping $S_P$ to a 7-brane that also wraps $S_P$, and 3-7' modes, which connect the instanton wrapping $S$ to a 7-brane wrapping a different 4-cycle which intersects $S$ along the matter curve $\Sigma$.

Our strategy for counting 3-7 modes is to start with the counting of 7-7 modes and then simply throw away the $SO(3,1)$ part.  We will later see that this is consistent with the usual perturbative counting in terms of D7-branes.  As we have seen, bifundamental 7-7 modes in the representation $R$ come from
\begin{equation}
S_+'\otimes\left[H^0(S,K_{S}\otimes V_R)\oplus H^1(S,V_R)\oplus H^2(S,K_{S}\otimes V_R)\right]\,.
\end{equation}
Dropping the $S_+'$ factor shows that 3-7 modes in the representation $R$ come from
\begin{equation}
H^0(S,K_{S}\otimes V_R)\oplus H^1(S,V_R)\oplus H^2(S,K_{S}\otimes V_R)\,.
\end{equation}

Counting the 3-7' modes is quite analogous.  In particular, 3-7' modes in the representation $R$ come from
\begin{equation}
H^0(\Sigma,K_{\Sigma}^{1/2}\otimes V_R)\,,
\end{equation}
while 7'-3 modes in the representation $\bar{R}$ come from
\begin{equation}
H^0(\Sigma,K_{\Sigma}^{1/2}\otimes V_R^{\ast})\,.
\end{equation}

\subsection{\textquotedblleft Perturbative\textquotedblright \ Counting of Fermion Zero Modes}\label{app:pertcount}

Let us now compare the above results for D3-instanton fermion zero modes to those obtained via a direct counting in the \textquotedblleft perturbative\textquotedblright \ type IIB language with D3's and D7's.  When a perturbative description is available, and assuming no orientifold planes are present near the D7-brane of interest, the 3-fold base $B$ is in fact Calabi-Yau.
\subsubsection{D3-Instanton: 3-3 Zero Modes}

We start by considering modes arising from strings which begin and end on the D3-instanton.  Fermion zero modes come from NN and DD directions which, for $3$-$3$ modes, are in fact all possible directions.  This means that the GSO projection simply leaves us with a positive chiral spinor with respect to $SO(10)$.

Letting $Adj(P)$ denote the adjoint bundle on the worldvolume $S$ of the D3-instanton, the D3-D3 fermion zero modes are sections of
\be
\ba
&S_+' \otimes \Biggl(H^{0}\left(S, K_{S}\otimes Adj(P)\right) + H^{2}\left(S, K_{S}\otimes Adj(P)\right)+
H^{1}\left(S,Adj(P)\right)\Biggr) \cr
\oplus\quad
& S_-' \otimes \Biggl(H^{0}\left(S,Adj(P)\right) + H^{2}\left(S,Adj(P)\right)+
H^{1}\left(S,K_{S}\otimes Adj(P)\right)\Biggr) \,,
\ea
\ee
where $S_+'$ ($S_-'$) denotes the
(anti-)chiral spin bundle in $\mathbb{R}^4.$
In the above, we have used the fact that although the del Pezzo surfaces do not admit a Spin structure, they nevertheless admit a Spin$^{c}$ structure, so that
the naive (anti-)chiral spin bundles on $S$ given by
\begin{equation}S_+=\left(\Omega^{0,0}\oplus\Omega^{0,2}\right)\otimes K_{S}^{1/2}\qquad
S_-=\Omega^{0,1}\otimes K_{S}^{1/2},\end{equation}
can combine with twists by $\mathcal{N}_{S/B}^{\pm 1/2}$ to form well-defined bundles.  In the explicit presentation
of zero modes given above, we have also used the isomorphism of line bundles
\be
\mathcal{N}_{S/B}=K_{S} \,,
\ee
which is valid when the 3-fold base of the F-theory compactification is Calabi-Yau.
See Section \ref{sec:generalities} for further discussion on this point.
Conjugating the anti-chiral spinor, it is now immediate that the fermion zero mode content
is in precise agreement with the result of subsection \ref{app:33}.


\subsubsection{D3-Instanton: 3-7 Zero Modes}

We now turn to the $3$-$7$ modes connecting the D3-instanton on $S$ to a 7-brane which also wraps $S$.  Fermions arise from NN and DD directions and hence come from directions along the base $B$. The GSO projection then leads to fermions of positive chirality with respect to the $SO(4)\times U(1)_R$ associated to $B$.  In particular, $3$-$7$ strings transform in the bifundamental representation $R$ and have positive chirality on $B$.  On the other hand, $7$-$3$ strings also have positive chirality with respect to $B$ but transform instead in the $\bar{R}$ representation.

The $3$-$7$ strings, which transform in the representation $R$, arise as sections of
\begin{equation}
\left[\left(S_+\otimes \mathcal{N}_{S/B}^{1/2}\right)\oplus\left(S_-\otimes \mathcal{N}_{S/B}^{-1/2}\right)\right]\otimes V_R
\end{equation}
and hence zero modes correspond to elements of
\begin{equation}
H^0(S,K_{S}\otimes V_R)\oplus H^1(S,V_R)\oplus H^2(S,K_{S}\otimes V_R) \,.
\end{equation}

For 7-3 strings, which transform in the representation $\bar{R}$, we have instead
\begin{equation}
\left[\left(S_+\otimes \mathcal{N}_{S/B}^{1/2}\right)\oplus\left(S_-\otimes \mathcal{N}_{S/B}^{-1/2}\right)\right]\otimes V_R^{\ast} \,.
\end{equation}
The corresponding zero modes are then elements of
\begin{equation}
H^0(S,K_{S}\otimes V_R^{\ast})\oplus H^1(S,V_R^{\ast})\oplus H^2(S,K_{S}\otimes V_R^{\ast})\,,
\end{equation}
which agrees with the result of subsection \ref{app:37}.


\subsubsection{D3-Instanton: 3-7' Zero Modes}

Finally, we study strings from the D3-instanton to a D7'-brane which intersect along the Riemann surface $\Sigma$.  Fermions again come from the NN and DD directions which in this case means that they come from directions along $\Sigma$.  The GSO projection then leads to fermions of positive chirality with respect to the corresponding $U(1)$ symmetry.

For 3-7' strings, which transform in the bifundamental representation $R$, the fermions are sections of
\begin{equation}
K_{\Sigma}^{1/2}\otimes V_R
\end{equation}
so that the zero modes are classified by
\begin{equation}
H^0(\Sigma,K_{\Sigma}^{1/2}\otimes V_R) \,.
\end{equation}

For 7'-3 strings, which transform in the bifundamental representation $\bar{R}$, the fermions are sections of
\begin{equation}
K_{\Sigma}^{1/2}\otimes V_R^{\ast}
\end{equation}
so that the zero modes are classified by
\begin{equation}
H^0(\Sigma,K_{\Sigma}^{1/2}\otimes V_R^{\ast})\,.
\end{equation}
This is in agreement with the results of subsection \ref{app:37}.


\section{Supersymmetric Instanton Bundles}
\label{app:GeneralSusyBundles}

One of the crucial results of the Polonyi-like model is that the only $X$-dependant contribution
from the 1-instanton sector is given by a linear term in the chiral superfield $X$.
This depends on classifying all possible solutions to the line bundle
conditions of equation (\ref{expl}).  In subsection \ref{app:arb},
we classify all such bundles for the del Pezzo 3 and 4 surfaces
for an arbitrary choice of polarization for the K\"ahler form.  For the
del Pezzo $M$ surfaces with $M > 4$, these conditions
appear to depend more sensitively on a particular ray in the K\"ahler cone.  Nevertheless,
in subsection \ref{app:largevolpol} we show that for a suitable large volume polarization
of the K\"ahler form, a similar result holds for $M > 4$ as well.  For completeness, in this same subsection
we also determine admissible supersymmetric line bundles which can contribute to a constant shift
in the superpotential.

\subsection{Solution for $dP_{3}$ and $dP_{4}$ for Arbitrary Polarization}\label{app:arb}

Here we solve (\ref{expl}) for  $S_1=dP_M$ with $M=3,4$
for a general choice of K\"ahler form $J.$
For $S_1=dP_3$  the condition $Q\ge 0$ becomes
$$
{7\over 2}\left(b_0+{m-3\over 7}\right)^2\le {8-3m^2-3m\over 7} \,.
$$
This condition can only be satisfied for $m = 0$ or $1$ since for $m>1$
the right side of this inequality is negative. Recall now that $b_0$ is an integer
so for $m=1$ the only solution is $b_0=0$ so that the D3-instanton line bundle is
\be \label{solviii}
V_{\text{inst}}=V_1\otimes \mathcal{O}(E_3-E_1-E_2) \,.
\ee
Now we recall that $V_{\text{inst}}$ must be trivializable in $B.$ We can choose the line bundle $V_1$
on $D7$ as
$V_1^{-1}=\mathcal{O}(E_3-E_1-E_2)$ so that  $V_{\text{inst}}=\mathcal{O}$ is trivial.

Now we consider $S_1=dP_4$. The condition $Q\ge 0$ may be written as
$$
{7\over 2}\left(b_0+{m+4x-3\over 7}\right)^2\le {(x+2m+1)^2-7y-7m^2-7m+7\over 7} \,.
$$
A necessary condition to satisfy the above inequality is
\be \label{nec}
\left(x+2m+1\right)^2\ge 7\left(y+m^2+m-1\right)\text{.}
\ee

Next recall the Cauchy-Schwarz inequality for vectors ${\vec v}$ and ${\vec w}$ in $\mathbb{R}^{2}$
which states
\be
({\vec v} \cdot {\vec w})^2 \le {\vec w}^2 {\vec v}^2 \,.
\ee
Using
$${\vec v}=(b_4,m+\half),\quad {\vec w}=(1,2) \,.$$
this becomes
\be \label{schw}
(x+2m+1)^2\le 5\left(y+m^2+m+{1\over 4}\right) \,.
\ee
Compatibility of (\ref{nec}) and (\ref{schw}) requires
$$\label{compi}2(b_4^2+m^2+m)\le 8+{1\over 4}$$
which can only be achieved when $m = 0$ or $1$.  Again restricting to the case of primary interest where $m = 1$,
this also implies $b_4=0$ or $b_4=1.$ Hence, there are two choices for the D3-instanton line bundle:
\be \label{solviv}
V_{\text{inst}}^{(j)}=V_1\otimes \mathcal{O}(E_j-E_1-E_2)\quad j=3,4 \,.
\ee
Again we recall that $V_{\text{inst}}$ must be trivializable in $B.$ We can choose the line bundle $V_1$
on $D7$ as
$V_1^{-1}=\mathcal{O}(E_3-E_1-E_2)$
then $V_{\text{inst}}^{(3)}=\mathcal{O}$ is obviously trivial, while
$V_{\text{inst}}^{(4)}=\mathcal{O}(E_4-E_3)$ can be trivializable for the appropriate choice of $B.$
For completeness, in the next subsection we will also determine all solutions with $b_{0} = m = 0$.

\subsection{Large Volume Polarization}\label{app:largevolpol}
In this subsection we establish a similar result for all $dP_{M}$ surfaces
when the K\"{a}hler class assumes the form:%
\begin{equation}
J=AH+\underset{i=1}{\overset{M}{\sum }}B_{i}E_{i}
\end{equation}%
where $A \gg 1$ and $B_{i}<0$ such that $\vert B_{i}\vert \sim O(1)$ for all $i$. \
Setting:%
\begin{equation}
c_{1}(\mathcal{L})=b_{0}H+\underset{i=1}{\overset{M}{\sum }}b_{i}E_{i}
\end{equation}%
as before, $\mathcal{L}$ defines a supersymmetric line bundle with respect
to this choice of polarization when:%
\begin{equation}
J\cdot c_{1}(\mathcal{L})=Ab_{0}-\underset{i=1}{\overset{M}{\sum }}%
b_{i}B_{i}=0\text{.}  \label{polcondL}
\end{equation}%
Our strategy for establishing a similar result to that found for the
particular case of $dP_{3}$ and $dP_{4}$ will be to bound the growth of $Q$.
\ The analysis naturally separates into the cases $b_{0}=0$ and $b_{0}\neq 0$.

As suggested by the $dP_{3}$ and $dP_{4}$ result, we now demonstrate that
no line bundle satisfying all of the zero mode conditions contributes when $%
b_{0}\neq 0$. \ Indeed, suppose to the contrary that such a line bundle can
contribute. \ The supersymmetric line bundle condition (\ref{polcondL})
implies that at least one of the $b_{i}$ grows rapidly enough that:%
\begin{equation}
O(\left\vert b_{i}\right\vert )\geq O(\left\vert Ab_{0}\right\vert )\gg
O(\left\vert b_{0}\right\vert )
\end{equation}%
as an order of magnitude estimate. \ We now show that for any choice of $%
i=1,...,M$, the resulting line bundle cannot satisfy (\ref{expl}).
\ The essential point in all cases is that in the definition of $Q$:%
\begin{equation}
Q=3-2(m+1)b_{0}+(b_{0}^{2}-2y)-m^{2}-2(x-1+2b_{0})^{2}\text{,}  \label{QDEF}
\end{equation}%
only $b_{0}^{2}$ and potentially $-2(m+1)b_{0}$ can provide a large positive
contribution to $Q$. \ We now show that the growth of the parameter $Q$ will
typically be large and negative, contradicting the fact that $Q$ must be
positive in order to achieve a solution.

First suppose that $i\geq 4$. \ This implies:%
\begin{equation}
O(y)\geq O(\left\vert Ab_{0}\right\vert ^{2})\gg O(b_{0}^{2})\text{.}
\label{OYOY}
\end{equation}%
Returning to the definition of $Q$, we note that because $y$ dominates over $%
b_{0}^{2}$, $2(m+1)b_{0}$ must be large and negative in order for $Q$ to be
positive. \ In particular, in order for this term to dominate over the
contribution from $2y$, we conclude that:%
\begin{equation}
O(\left\vert mb_{0}\right\vert )\geq O(y)\gg O(b_{0}^{2})
\end{equation}%
or:%
\begin{equation}
O(m^{2})\gg O(\left\vert mb_{0}\right\vert )\text{.}
\end{equation}%
Returning to equation (\ref{QDEF}), it follows that the negative
contribution $-m^{2}$ dominates over the term $-2(m+1)b_{0}$. \ Hence, for
sufficiently large $A\gg 1$, $Q$ is always negative so that no solution
exists.

As an intermediate case, next suppose that $i=3$. \ Returning to the
definition of $b_{3}$, we note that:%
\begin{equation}
b_{3}=1-2b_{0}-x\text{.}
\end{equation}%
In order for $O(\left\vert b_{3}\right\vert )\gg O(\left\vert
b_{0}\right\vert )$, we must require $O(\left\vert x\right\vert
)=O(\left\vert b_{3}\right\vert )\geq O(\left\vert Ab_{0}\right\vert )$. \
This in turn implies that at least one of the summands of $x$ must grow as
rapidly as $\left\vert b_{3}\right\vert $, which falls under the analysis
below (\ref{OYOY}).

Next consider the cases $i=1$ and $i=2$. \ Recall that $b_{1}$
and $b_{2}$ satisfy the relations:%
\begin{eqnarray}
b_{1} &=&-\frac{1}{2}\left( (b_{0}+m+1)\mp \sqrt{Q}\right)  \\
b_{2} &=&-\frac{1}{2}\left( (b_{0}+m+1)\pm \sqrt{Q}\right) \text{.}
\end{eqnarray}%
In both cases, we therefore conclude that either $O(\left\vert m\right\vert
)\geq O(\left\vert b_{i}\right\vert )$ or $O\left( \left\vert \sqrt{Q}%
\right\vert \right) \geq O(\left\vert b_{i}\right\vert )$. \ First suppose $%
O(\left\vert m\right\vert )\geq O(\left\vert b_{i}\right\vert )\geq
O(\left\vert Ab_{0}\right\vert )$. \ In this case, we have:%
\begin{equation}
O(m^{2})\gg O(\left\vert mb_{0}\right\vert )\gg O(b_{0}^{2})\text{.}
\end{equation}%
By inspection of $Q$, we therefore conclude that $-m^{2}$ dominates over the
two potentially large positive contributions $b_{0}^{2}$ and $-2(m+1)b_{0}$.
\ Hence, we again find that $Q$ is always negative in this case.

Finally, suppose that $O\left( \left\vert \sqrt{Q}\right\vert \right) \geq
O(\left\vert b_{i}\right\vert )\geq O(\left\vert Ab_{0}\right\vert )$. \ If
$Q$ is large and positive, then:%
\begin{equation}
O(\left\vert mb_{0}\right\vert )\geq O(\left\vert Ab_{0}\right\vert ^{2})
\end{equation}%
or:%
\begin{equation}
O(\left\vert m\right\vert )\geq O(\left\vert A^{2}b_{0}\right\vert )\gg
O(\left\vert b_{0}\right\vert )\text{.}  \label{lastmbound}
\end{equation}%
In particular, this implies $O(\left\vert m\right\vert ^{2})\gg O(\left\vert
mb_{0}\right\vert )$ so that as before, $Q$ is negative when $A$ is
sufficiently large. \ In all cases, it thus follows that we achieve a
contradiction when $b_{0}\neq 0$.

Next consider the case $b_{0}=0$. \ Here, positivity of $Q$ again implies
stringent conditions on the admissible line bundles. \ Indeed, when $b_{0}=0$%
, we have:%
\begin{equation}
Q(b_{0}=0)=3-2y-m^{2}-2(x-1)^{2}\text{.}
\end{equation}%
In particular, we note that positivity of $Q$ imposes the conditions:%
\begin{equation}
0\leq m\leq 1
\end{equation}%
and $x=y=0$, or $x=y=1$. \ The only available values are therefore given by:%
\begin{eqnarray}
b_{0} &=&m=0\text{, }x=y=0 \\
b_{0} &=&m=0\text{, }x=y=1 \\
b_{0} &=&x=y=0\text{, }m=1 \\
b_{0} &=&0\text{, }x=y=m=1\text{.}
\end{eqnarray}%
In the $m = 0$ sector, this implies $Q = 1$, while in the $m = 1$ sector, this implies $Q = 0$.  This yields the following supersymmetric line bundle configurations which can contribute to the Polonyi-like model:
\begin{eqnarray}
m &=& 0: \mathcal{L} = \mathcal{O}(E_{p} - E_{j}) \\
m &=& 1: \mathcal{L} = \mathcal{O}(E_{p}-E_{1}-E_{2}) \text{,}
\end{eqnarray}%
for $j=1,2$ and $p = 3,...,M$. Assuming that the volume modulus of the del Pezzo surface has been stabilized at a sufficiently high energy scale, the $m = 0$
sector characterizes relatively unimportant constant shifts in the 4d superpotential.

Note that at most one of the m's can
contribute. Indeed, we have two candidate ${\mathcal{L}}$'s for the two sectors:
$\mathcal{L}_{m=0}$ and $\mathcal{L}_{m=1}$.
Now, letting $V$ denote the corresponding bundle on the $D3$, we have:
$$\mathcal{L}_{m=0} \otimes \mathcal{L}_{m=1}^{-1} =V_{m=0} \otimes V_{m=1}^{-1}$$
So, since the product of $V$'s is not trivializable, it follows that at least
one of the fluxes through the D3 is not trivializable.
We prefer to choose $V_1^{-1}=\mathcal{O}(E_3-E_1-E_2)$ 
then $V_{m=0}$ will be non-trivializable  and we get no constant shift of superpotential.
Meanwhile, in $m=1$ sector we have either trivial instanton bundle $V_{m=1,\, p=3}=\mathcal{O}$ or
instanton bundle $V_{m=1, \, p}=\mathcal{O}(E_p-E_3)$ for $p=4,\ldots,M$  trivializable for some choice of $B.$



\newpage

\bibliographystyle{JHEP}
\renewcommand{\refname}{Bibliography}

\providecommand{\href}[2]{#2}\begingroup\raggedright\endgroup


\end{document}